\documentclass[prd, nofootinbib, floatfix, notitlepage]{revtex4-1}

\usepackage{amsmath,amsfonts,amssymb}
\usepackage{mathrsfs}
\usepackage{graphicx}
\usepackage[english]{babel} 
\usepackage{hyperref} 
\usepackage{color}
\usepackage{bbold}

\newcommand{\be}{\begin{equation}}
\newcommand{\ee}{\end{equation}}
\newcommand{\bes}{\begin{equation*}}
\newcommand{\ees}{\end{equation*}}

  \usepackage[utf8]{inputenc} 

\begin{document}

\title{Sensitivity to a Frequency-Dependent Circular Polarization in an Isotropic Stochastic Gravitational Wave Background}
\author{Tristan L.~Smith$^1$ and Robert Caldwell$^2$ }

\begin{abstract}

We calculate the sensitivity to a circular polarization of an isotropic stochastic gravitational wave background (ISGWB) as a function of frequency for ground- and space-based interferometers and observations of the cosmic microwave background. The origin of a circularly polarized ISGWB may be due to exotic primordial physics (i.e., parity violation in the early universe) and may be strongly frequency dependent. We present calculations within a coherent framework which clarifies the basic requirements for sensitivity to circular polarization, in distinction from previous work which focused on each of these techniques separately.  We find that the addition of an interferometer with the sensitivity of the Einstein Telescope in the southern hemisphere improves the sensitivity of the ground-based network to circular polarization by about a factor of two.    The sensitivity curves presented in this paper make clear that the wide range in frequencies of current and planned observations ($10^{-18}\ {\rm Hz} \lesssim f \lesssim 100\ {\rm Hz}$) will be critical to determining the physics that underlies any positive detection of circular polarization in the ISGWB. We also identify a desert in circular polarization sensitivity for frequencies between $10^{-15}\ {\rm Hz} \lesssim f  \lesssim 10^{-3}\ {\rm Hz}$, given the inability for pulsar timing arrays and indirect-detection methods to distinguish the gravitational wave polarization.
\end{abstract}

\affiliation{$^1$Department of Physics \& Astronomy, Swarthmore College, Swarthmore, PA 19081 USA\\ $^2$Department of Physics \& Astronomy, Dartmouth College, Hanover, NH 03755 USA }
 \date{\today}

\maketitle

\section{Introduction}

Starting with the first stargazers, our knowledge of the heavens has come in the form of electromagnetic waves.  The intensity and polarization of these massless messengers have been shown to contain a wealth of information about the physics and astrophysics of distant objects and the conditions along the line of sight, extending to the earliest moments after the primordial universe became transparent.  Now the same is coming true for gravitational waves \cite{Abbott:2016blz}. 

In this paper we consider the frequency-dependent sensitivity of the most common gravitational wave detection techniques to a net circular polarization of an isotropic stochastic gravitational wave background (ISGWB).  Since gravitational waves have two polarizations, any stochastic gravitational wave background can be expanded in terms of the standard four Stokes parameters: $I$, $Q$, $U$, and $V$.  However, given the spin-2 nature of gravitational waves, the $Q$ and $U$ linear polarizations are only non-zero for anisotropic backgrounds (with the first non-zero contribution at the quadrupole).  On the other hand both $I$ and $V$ are scalar quantities, and as such, may be non-zero for isotropic backgrounds.  Since most stochastic gravitational wave backgrounds are predicted to be nearly isotropic we only consider the sensitivity of the most common techniques to the intensity, $I$, and level of circular polarization, $V$. 

The detection of a non-zero circularly-polarized ISGWB would indicate new fundamental physics \cite{PhysRev.106.388}. Leading examples consist of inflationary models in which the inflaton couples to the parity-odd Chern-Simons scalar of a U(1) vector field, as in Refs.~\cite{Anber:2009ua,Anber:2012du}, or the inflaton couples similarly to a non-Abelian SU(2) gauge field, as in chromo-natural inflation \cite{Adshead:2012kp,Adshead:2013qp} or gauge-flation \cite{Maleknejad:2011jw,Namba:2013kia} and its variants \cite{Maleknejad:2012fw,Cai:2016ihp,Dimastrogiovanni:2016fuu,Adshead:2016}.  Through different mechanisms, these scenarios all generate a primordial spectrum of gravitational waves with a scale-dependent chiral asymmetry, whereby the spectra for left- and right-circular polarizations differ. An inflaton that couples directly to the gravitational Chern-Simons scalar  \cite{Alexander:2004us,Alexander:2009tp} and other quantum gravity schemes \cite{Takahashi:2009wc,Contaldi:2008yz} will also generate an asymmetry. In the case of chromo-natural and gauge-flation, the spectrum is chirally-symmetric below a cutoff scale typically close to the present-day horizon radius, and chirally-asymmetric with a blue tilt on smaller scales. This opens the possibility of detection across a wide range of frequencies by the cosmic microwave background (CMB), satellite, and ground-based detectors. 

Numerous gravitational wave observatories are on line or in planning stages. LIGO and VIRGO are already taking data; Pulsar Timing arrays (PTAs) may expect to see a signal in the near future; LIGO India and KAGRA are under development; and a global network of ground-based interferometers has been proposed, under the name Einstein Telescope. Technology is already developing for space-based observatories such as the Evolved Laser Interferometer Space Antenna (eLISA) \cite{eLISA} (recently re-named LISA) and the Big Bang Observer (BBO) \cite{Crowder:2005nr}. These independent but complementary observatories are sensitive to different frequencies. On the largest scales (i.e., frequencies of $f \simeq 10^{-18}\ {\rm Hz}$) we may detect gravitational waves through their effects on the CMB.  Observations of both the intensity (i.e., temperature) and linear polarization of the CMB give information about the properties of a stochastic gravitational wave background on scales equal to the size of the observable universe. The autocorrelation of the temperature and B-mode polarization provides an estimate of the intensity of a possible ISGWB whereas the cross-correlation of the temperature and B-mode polarization as well as the cross-correlation of the E and B-mode polarization  provide estimates of the level of net circular polarization.    Pulsar timing arrays (PTAs) are most sensitive to gravitational waves at frequencies $f \simeq 10^{-9}\ {\rm Hz}$.  However, because of the effective geometry of the detector -- the fact that we measure each pulse time of arrival at the Earth -- PTAs are not sensitive to the circular polarization of an ISGWB.  At moderate frequencies, $f \simeq 1\ {\rm Hz}$, space-based laser interferometers can be made to be sensitive to the circular polarization of an ISGWB by correlating the signals recorded by two independent observatories lying in different planes.  Finally at high frequencies ($f \simeq 100\ {\rm Hz}$) the correlation between signals of ground-based laser interferometers are already sensitive to the circular polarization of the ISGWB. 

Previous work has considered the sensitivity to the circular polarization of the ISGWB for gound-based and space-based gravitational wave observatories.  Refs.~\cite{Seto:2006hf,Seto:2006dz,Seto:2007tn,Seto:2008sr,Seto:2008sr2} consider the sensitivity of ground-based and space-based interferometers and Ref.~\cite{Gluscevic:2010vv} considers the sensitivity of measurements of CMB polarization.  Much of this is included in the exhaustive review in Ref.~\cite{Romano:2016dpx}. We extend this work in several ways. First, this paper presents the sensitivity to circular polarization in a consistent framework for each observatory.  This allows us to gain a clearer intuition for what type of observatory will provide useful information on the circular polarization as well as provide formulae which can be used to calculate the sensitivity of future observatories.  Second, while previous work calculated the sensitivity to a flat spectrum (i.e. $\Omega_{\rm gw} = {\rm constant}$), we present the full sensitivity curves, which determines the frequency range associated with each observatory and allows a comparison of the sensitivity to non-flat spectra.  Finally, we consider the sensitivity of several observatories (such as the Einstein Telescope) which were not included in previous work. 

This paper is organized as follows: In Sec.~II we discuss the basic physics of a gravitational wave detector, present a calculation of the optimal signal to noise, discuss the properties of the ISGWB, and present the method we use to calculate the sensitivity curves.  In Sec.~III we calculate the sensitivity curves for space-based observatories.  In Sec.~IV we calculate the sensitivity curves for a network of ground-based observatories.  In Sec.~V we calculate the sensitivity curves for observations of the CMB.  In Sec.~VI we present our conclusions. 


\section{Detecting the gravitational wave background}

We consider the detection of gravitational waves by an interferometer, generalizing the design concept of LIGO \cite{33318}: the relative shift in the phase of light beams traveling between test masses in the two arms of an interferometer is used to detect the presence of a gravitational wave. The effect of a gravitational wave on this relative phase can be simply calculated from knowledge of the motion of null geodesics in a nearly flat spacetime \cite{Finn:2008np,Cornish:2009rt}.  The following calculation closely follows the calculation presented in Ref.~\cite{Cornish:2001bb}. To define the gravitational wave transfer function we expand the gravitational wave background in plane waves:
\begin{equation}
h_{ab}(\vec x,t) = \int_{-\infty}^{\infty}df  \int d^2\hat n \, \sum_P \, \tilde h_P(f,\hat n) e^P_{ab}(\hat{n})e^{i2 \pi f(t - \hat n \cdot \vec x)},
\end{equation}
where $e^P_{ab}$ is the polarization tensor.  For a $P=+,\,\times$ polarized plane wave propagating in the $\hat n = (\cos\phi \, \sin\theta, \sin\phi\, \sin\theta, \cos\theta)$ direction, the polarization tensors may be written
\begin{eqnarray}
e^{+}_{ab}(\hat n) &=& \hat{ \mathbb{m}}_a \hat{\mathbb{m}}_b - \hat{\mathbb{n}}_a \hat{\mathbb{n}}_b \\
e^{\times}_{ab}(\hat n) &=& \hat{ \mathbb{m}}_a \hat{ \mathbb{n}}_b + \hat{ \mathbb{n}}_a\hat{ \mathbb{m}}_b \\
\hat{ \mathbb{m}} &\equiv& (\sin\phi, -\cos\phi, 0) \\
\hat{ \mathbb{n}} &\equiv& (\cos\phi\, \cos\theta, \sin\phi\, \cos\theta, -\sin\theta)
\end{eqnarray}
so that $e^{P}_{ab}(\hat n) e^{P',ab}(\hat n) = 2 \delta_{P P'}$ and $\hat{ \mathbb{m}},\,\hat{ \mathbb{n}}$ are Newman-Penrose vectors.  These polarization tensors can also be written in a circular polarization basis, 
\begin{eqnarray}
e^{R}_{ab}(\hat n) &=& \frac{e^{+}_{ab}(\hat n) + i e^{\times}_{ab}(\hat n)}{\sqrt{2}},\\
e^{L}_{ab}(\hat n) &=& \frac{e^{+}_{ab}(\hat n) - i e^{\times}_{ab}(\hat n)}{\sqrt{2}}.
\end{eqnarray}
We can use this expansion to express the phase accumulated along a single arm of the interferometer as
\begin{equation}
\varphi_{1 2}(t_1) =\varphi_0\left[1 + \int_{-\infty}^{\infty} df \int d^2 \hat{n}\  \sum_P \tilde h_P(f,\hat n) e^P_{ab}(\hat{n})e^{i 2\pi f(t_1- \hat{n} \cdot \vec x_1)} \mathcal{D}^{ab}(\hat{\ell}_{12} \cdot \hat{n},f)\right],
\end{equation}
where $t_1$ is the time at which the light left mass 1, $\vec x_1$ is the location of mass 1, $\vec x_1 + L \hat{\ell}_{12}$ is the location of mass 2, and
the single-arm transfer function is given by 
\begin{equation}
\mathcal{D}^{ab}(\hat{\ell} \cdot \hat{n},f) \equiv \frac{1}{2} \hat{\ell}^a \hat{\ell}^b \mathcal{M}(\hat{\ell} \cdot \hat{n},f),
\end{equation}
where 
\begin{equation}
\mathcal{M}(\hat{\ell} \cdot \hat{n}, f) \equiv {\rm sinc}\left[\frac{f}{2f_*} (1-\hat{\ell} \cdot \hat{n})\right] e^{i[f/(2f_*)(1-\hat{\ell} \cdot \hat{n})]} = \frac{i f_*}{f} \frac{e^{i f/f_*(1-\hat{\ell} \cdot 
\hat{n})}-1}{1-\hat{\ell} \cdot \hat{n}}
\end{equation}
and $f_* \equiv (2\pi L)^{-1}$ is the characteristic frequency scale of the detector. Note that for a single arm the response is approximately equal to 1 for $f\ll f_*$ and decreases as $1/f$ for $f \gg f_*$.  
\begin{figure}[h!]
\begin{center}
\resizebox{!}{7cm}{\includegraphics{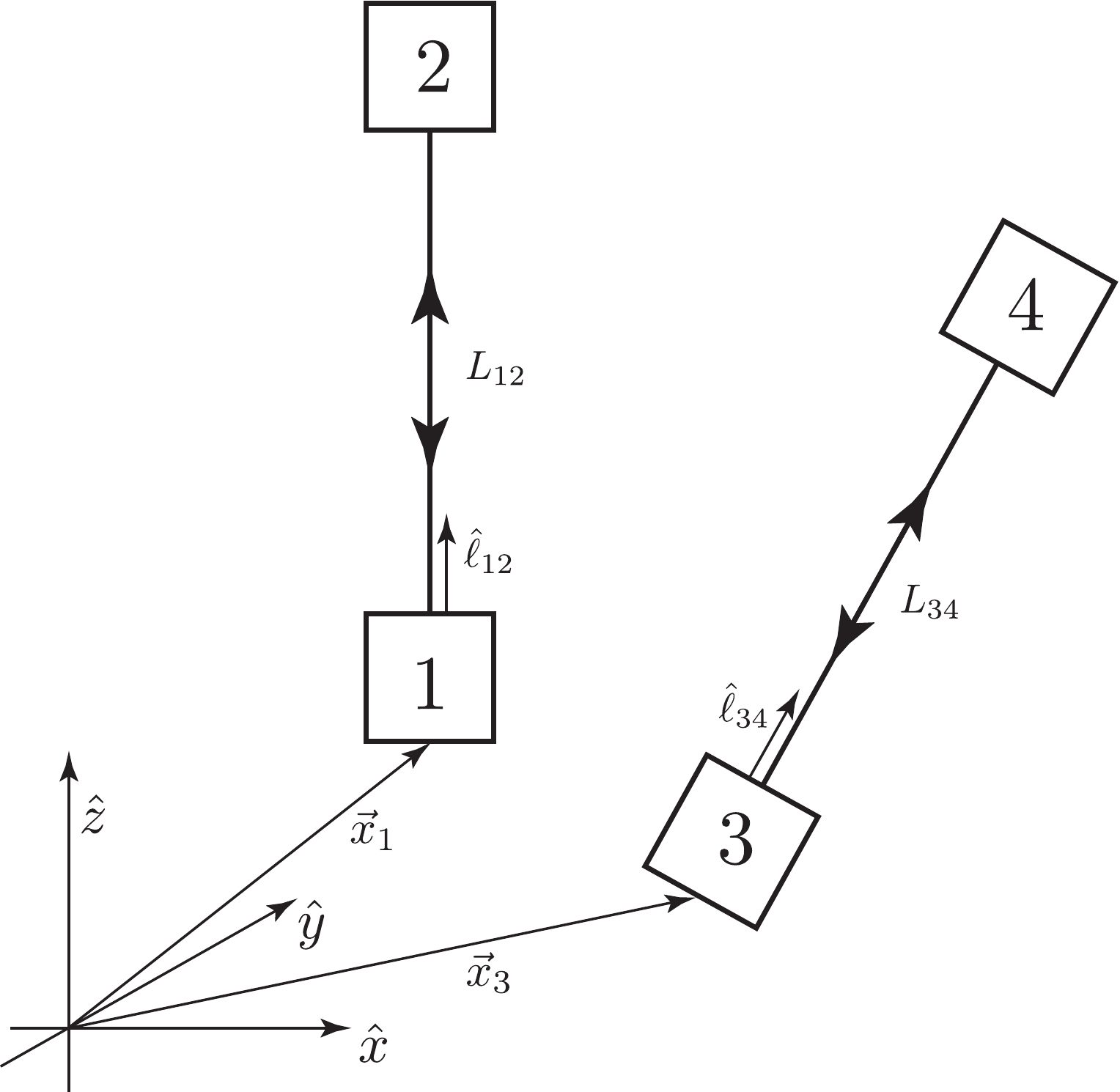}}
\caption{A schematic figure showing the geometry of the four detectors we are considering in this calculation.}
\end{center}
\end{figure}

In order to get a sense of how the sensitivity curve is calculated let us build up our detector starting with a single arm. The change in phase of the light beam as it passes from one end of the arm and then back again is
\begin{eqnarray}
s_1(t) &\equiv&  \Delta \varphi_{1 2}(t-2L) + \Delta \varphi_{21}(t-L)+n_1(t),
\end{eqnarray}
where $\Delta \varphi(t) \equiv [\varphi(t) - \varphi_0]/\varphi_0$ and $n_1(t)$ is a noise term. It is useful to consider the Fourier transform of the signal, 
\begin{equation}
\tilde s_1(f) = \Delta \tilde{\varphi}_{12}(f) e^{-i 2 \pi f (2L)} +\Delta \tilde{\varphi}_{21}(f) e^{-i 2 \pi f L}+ \tilde n_1(f), 
\end{equation}
where $\tilde A(f) \equiv \int_{-T/2}^{T/2} dt  A(t) e^{-i2 \pi f t}$ and 
\begin{equation}
\Delta\tilde{ \varphi}_{i  j}(f) = \int_{-\infty}^{\infty} df' \delta_T(f-f')\int d^2 \hat{n} e^{-i 2\pi f' \hat{n} \cdot \vec x_i}\tilde h_P(f',\hat{n} )e^P_{ab}(\hat n) \mathcal{D}^{ab}(\hat{\ell}_{ij} \cdot \hat{n},f'),
\end{equation}
where $\delta_T(f-f') \equiv T {\rm sinc}[(f-f') \pi T]$; we have $\lim_{T\rightarrow \infty} \delta_T(f-f') = \delta(f-f')$. To measure the stochastic background we need to correlate this signal with one from another arm:
\begin{eqnarray}
s_3(t) &\equiv&  \Delta \varphi_{3 4}(t-2L) + \Delta \varphi_{43}(t-L)+n_3(t).
\end{eqnarray}
The correlation of any two of the phase differences from these measurements is 
\begin{eqnarray}
\langle \Delta\tilde{ \varphi}_{i  j}(f) \Delta\tilde{ \varphi}_{k  l}^*(f')\rangle &=&  \frac{1}{2} \int_{-\infty}^{\infty} df''  \delta_T(f-f'') \delta_T(f'-f'') S_h^{PP'}(f'')\mathcal{R}^{ij,kl}_{PP'}(f''),\\
\mathcal{R}^{ij,kl}_{PP'}(f'')&\equiv&\int \frac{d^2 \hat{n}}{4\pi} e^{i 2 \pi f''\hat n\cdot (\vec x_k-\vec x_i)} \mathcal{D}_{ab}(\hat{\ell}_{ij} \cdot \hat{n},f'')e^{ab}_P(\hat n)\mathcal{D}^*_{cd}(\hat{\ell}_{kl} \cdot \hat{n},f'') e^{cd}_{P'}(\hat n),
\end{eqnarray}
where ${\cal R}$ is the response function, and we assume the correlation between the gravitational wave Fourier modes takes the form 
\begin{equation}
\langle \tilde{h}_P(f,\hat{n}) \tilde{h}^*_{P'}(f',\hat{n}') = \frac{1}{2} S_h^{PP'}(f) \delta(f-f') \frac{\delta^{(2)}(\hat n - \hat n')}{4\pi}. 
\end{equation}
Since the observatories operate over a time-scale of several years, $T\sim 10^8\ {\rm s}$, and at frequencies $10^{-5}\ {\rm Hz} \lesssim f \lesssim 10^3\ {\rm Hz}$ we always have $fT \gg 1$ and $\delta_T(f)$ can be well-approximated as a Dirac delta function so that 
\begin{equation}
\langle \Delta\tilde{ \varphi}_{i  j}(f) \Delta\tilde{ \varphi}_{k  l}^*(f')\rangle \simeq \frac{1}{2}  S_h^{PP'}(f)\mathcal{R}^{ij,kl}_{PP'}(f) \delta_T(f-f') \equiv \frac{1}{2}  S_s(f)\delta_T(f-f'),
\end{equation}
where $S_s(f)$ is the signal power spectrum. 

\subsection{The optimal signal to noise ratio} 
\label{sec:opSNR}

Imagine a collection of $N$ signals, $\{\tilde{s}_i(f)\}$, from which we can construct the frequency-dependent estimator as 
\begin{equation}
\hat{\mathcal{C}}(f,f') \equiv \frac{1}{2} W^{ij}(f,f')\tilde{s}_i(f)\tilde{s}^*_j(f'),
\end{equation}
where the weight matrix is symmetric, $W^{ij}(f,f')= W^{*ji}(f',f)$, an even function in $f$ and $f'$, and zero along the diagonal, $W^{ii}(f,f') = 0$. 
From this estimator we can construct the frequency-integrated estimator 
\begin{equation}
\hat{\mathcal{C}} \equiv \frac{1}{2} \int_{-\infty}^{\infty} df df' W^{ij}(f,f')\tilde{s}_i(f)\tilde{s}^*_j(f'). 
\end{equation}
The expectation value of the estimator is 
\begin{equation}
\langle \hat{\mathcal{C}}\rangle = \frac{1}{4}\int_{-\infty}^{\infty} df df' \delta_T(f-f')W^{ij}(f,f')S_{s,ij}(f),
\end{equation}
where, for example, with $i=1$ and $j=3$ we have
\begin{equation}
S_{s,13}(f) = S_h^{PP'}(f) \left[ \mathcal{R}^{12,34}_{PP'}(f)+\mathcal{R}^{21,43}_{PP'}(f)+e^{-i2\pi fL}\mathcal{R}^{12,43}_{PP'}(f)+e^{i2\pi fL}\mathcal{R}^{21,34}_{PP'}(f)\right].
\end{equation}
Assuming the noise power spectrum takes the form $\langle \tilde{n}_i(f) \tilde{n}^*_j(f')\rangle = \frac{1}{2} S_{n}^{(i)}(f)\delta_{ij} \delta_T(f-f') $ and that we are dealing with weak, noise dominated signals so that $S^{(i)}_{n}(f) \gg S_{s,ij}(f)$, then the variance of this estimator is given by 
\begin{equation}
\sigma_{\hat{\mathcal{C}}}^2 \simeq \langle \hat{\mathcal{C}}^2\rangle = \frac{1}{8}\int_{-\infty}^{\infty} df df' W^{ij}(f,f')S_{n}^{(i)}(f)W^{*}_{ij}(f,f')S^{j}_{n}(f').
\end{equation}
The signal-to-noise ratio (SNR) of this measurement is then given by 
\begin{equation}
{\rm SNR} \simeq \frac{1}{\sqrt{2}}\frac{\int_{-\infty}^{\infty} df df' \delta_T(f-f')W^{ij}(f,f')S_{s,ij}(f)}{\sqrt{\int_{-\infty}^{\infty} df df' W^{ij}(f,f')S_{n}^{(i)}(f)W^{*}_{ij}(f,f')S_{n}^{(j)}(f')}}.
\end{equation}
To determine what filter function $W_{ij}(f,f')$ will maximize the SNR, we introduce a noise-weighted inner product 
\begin{equation}
(A_{ij},B_{ij}) \equiv \int_{-\infty}^{\infty} df df' A^{ij}(f,f') B^{*}_{ij}(f,f') S^{(i)}_{n}(f) S^{(j)}_{n}(f').
\end{equation}
With this the SNR can be written as 
\begin{equation}
{\rm SNR} =\frac{1}{\sqrt{2}}\frac{\left(W^{ij}(f,f'), \frac{S_{s,ij}(f)\delta_T(f-f')}{S^{(i)}_{n}(f) S^{(j)}_{n}(f')}\right)}{\sqrt{(W^{ij}(f,f'), W^*_{ij}(f,f'))}}.
\end{equation}
It is clear that this will be maximized if $W_{ij}(f,f') = \lambda S_{s,ij}(f) \delta_T(f-f')/ [S^{(i)}_{n}(f) S^{(j)}_{n}(f')]$, where $\lambda$ is some normalization.  With this choice, the optimal SNR is given by 
\begin{equation}
{\rm SNR} = \left[T\sum_{i<j}\int_{-\infty}^\infty df \frac{S^2_{s,ij}(f)}{S^{(i)}_{n}(f) S^{(j)}_{n}(f)}\right]^{1/2}.
\label{eq:maxSNR}
\end{equation}

\subsection{Stochastic Background}

Consider an ISGWB with zero mean.  If there is a net polarization then the variance is given by 
\begin{eqnarray}
\left(\begin{array}{cc}\langle h_+^*(f,\hat{n}) h_+(f',\hat{n}')\rangle & \langle h_+^*(f,\hat{n}) h_{\times}(f',\hat{n}')\rangle \\ \langle h_\times^*(f,\hat{n}) h_+(f',\hat{n}')\rangle & \langle h_\times^*(f,\hat{n}) h_{\times}(f',\hat{n}')\rangle\end{array}\right) &=& \frac{1}{2} \delta(f-f') \frac{\delta^{(2)}(\hat{n} - \hat{n'})}{4\pi} \left(\begin{array}{cc}I +Q & U+iV \\U-iV & I-Q\end{array}\right),\\
&=& \frac{1}{2} \delta^{(3)}(\vec k-\vec k') \left(\begin{array}{cc}I +Q & U+iV \\U-iV & I-Q\end{array}\right).
\end{eqnarray}
The overall intensity, $I$, and circular polarization, $V$, are scalar quantities, and hence can be measured through the monopole of the stochastic background; the $Q$ and $U$ are spin-4 quantities and hence do not contribute to an isotropic, stochastic, background. Since we are considering an isotropic background, for the rest of this discussion we will take $Q=U=0$. This leads to the result 
\begin{equation}
 \langle \Delta \tilde \varphi_{i j}(f) \Delta \tilde \varphi^*_{k l}(f')\rangle = \frac{1}{2} \left[ \mathcal{R}^I_{ij,kl}(f) I(f)+ \mathcal{R}^V_{ij,kl}(f) V(f)\right]\delta(f-f'),
\end{equation}
where 
\begin{eqnarray}
\mathcal{R}^I_{ij,kl}(f) &=& \frac{1}{4\pi}\int d^2 \hat{n} \left[F^+_{ij}(\hat n,f)F^{*+}_{kl}(\hat n,f)+F^\times_{ij}(\hat n,f)F^{*\times}_{kl}(\hat n,f)\right], \\
\mathcal{R}^V_{ij,kl}(f) &=& \frac{i}{4\pi}\int d^2 \hat{n} \left[F^+_{ij}(\hat n,f) F^{*\times}_{kl}(\hat n,f) - F^{\times}_{ij} (\hat n,f)F^{*+}_{kl}(\hat n,f)\right],
\end{eqnarray}
and where
\begin{equation}
F^P_{ij}(\hat n,f) \equiv  e^{-i 2\pi f\hat{n} \cdot \vec x_i}e_{ab}^P(\hat n) \mathcal{D}^{ab}(\hat{\ell}_{ij} \cdot \hat{n},f).
\end{equation}

Without loss of generality we can place $\vec x_i$ at the origin of our coordinate system, $\vec x_k$ along the $z$-axis, and $\hat{\ell}_{ij}$ in the $x-z$ plane so that $\vec x_k = D \hat{z}$ and $\hat{\ell}_{ij} = \cos \alpha \hat x + \sin \alpha \hat z$.  Most gravitational wave observatories, such as LIGO, LISA, and PTAs, effectively have only three masses which, as a result, are necessarily co-planar. The same is true for most designs for futuristic space-based gravitational wave observatories such as the Big Bang Observer (BBO) and the Decihertz Gravitational Wave Observatory (DECIGO), each of which have advanced stages with six masses \cite{Crowder:2005nr, kawamura2006japanese}.  In the case of a co-planar observatory, we can also write $\hat{\ell}_{kl} = \cos \beta \hat x + \sin \beta \hat z$.  It is straight forward to show that in this case if we reflect about the plane of the observatory (i.e., $\phi \rightarrow - \phi$) we have $F^+_{ij}(\hat n,f) \rightarrow F^+_{ij}(\hat n,f)$ and $F^\times_{ij}(\hat n,f) \rightarrow -F^\times_{ij}(\hat n,f)$ so that $\mathcal{R}^V_{ij,kl}(f)  = 0$. This result is not surprising: for a planar observatory a right-handed gravitational wave coming from `above' is indistinguishable from a left-handed gravitational wave traveling from `below'. Therefore only those observatories constructed from masses which are non-co-planar will be sensitive to the circular polarization of an isotropic stochastic gravitational wave background.  This means that PTAs are only sensitive to the intensity of the ISGWB. 

\subsection{Sensitivity curve}

With an expression for the SNR we write the total SNR as the sum of the sliding integral:
\begin{equation}
{\rm SNR}^2 =  \sum_{f_i} 2T \int_{f_i-\Delta f/2}^{f_i+\Delta f/2}\frac{S^2_{s}(f)}{S_{n,1}(f) S_{n,3}(f)}df \equiv \sum_{f_i} {\rm SNR}^2(f_i).
\end{equation}
Writing $S_{s}(f) \equiv S_h^{PP'}(f) \mathcal{R}_{PP'}(f) = (3 H_0^2)/(4\pi^2) f^{-3} \Omega^{PP'}_{\rm gw}(f)\mathcal{R}_{PP'}(f)$ \cite{Maggiore:1999vm} we can write the minimum-detectable  gravitational wave background within a bandwidth $\Delta f$ as (i.e., a sensitivity curve) \cite{Cornish:2001bb}
\begin{equation}
\Omega^{PP',{\rm min}}_{\rm gw}(f_i) \simeq {\rm SNR}_0 \left[2 T  \int_{f_i-\Delta f/2}^{f_i+\Delta f/2} \left(\frac{3 H_0^2}{4\pi^2}\right)^2 \frac{ \mathcal{R}_{PP'}(f)}{f^{6}S_{n,1}(f) S_{n,3}(f)} df\right]^{-1/2}.
\end{equation}
For all of the sensitivity curves we take $\Delta f = 0.05 f_i$, $H_0= 72$ km/s/Mpc \cite{Riess:2016jrr}, and $T=10$ years. 
An ISGWB spectrum that exceeds this sensitivity curve will be detectable with an SNR $\gtrsim {\rm SNR}_0$. 
 
\section{Space-based interferometers}

\begin{figure}[h!]
\begin{center}
\resizebox{!}{7cm}{\includegraphics{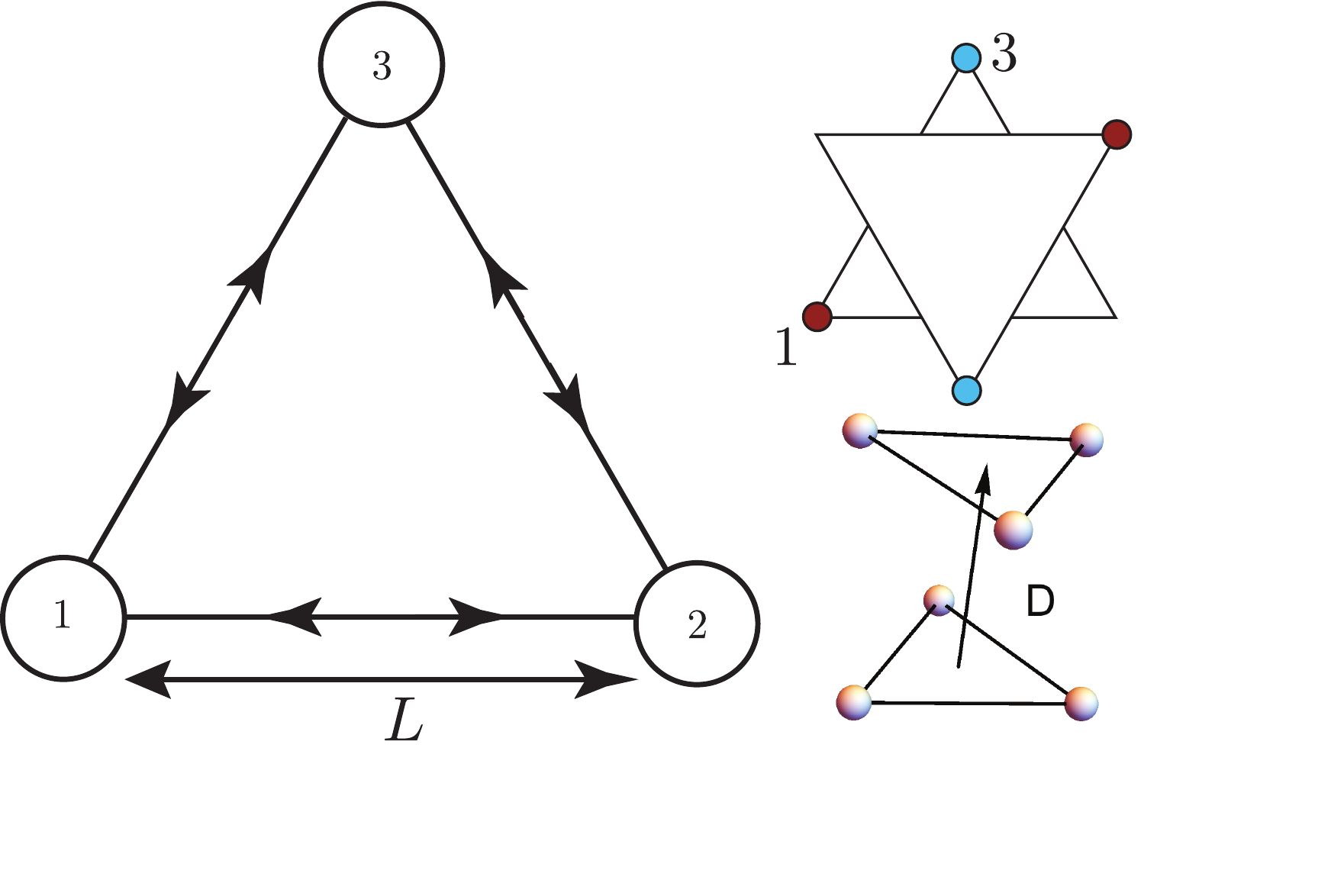}}
\caption{Two equal arm Michelson interferometers rotated by 180 degrees and separated by a distance $D$.}
\label{fig:space-based}
\end{center}
\end{figure}

A space-based interferometer that is sensitive to the circular polarization of the ISGWB can consist of two equilateral triangles with barycenters separated by a distance $D$ (see Fig.~\ref{fig:space-based}) \cite{Seto:2006hf,Seto:2006dz}.  In order to calculate the sensitivity for the nominal design for various space-based gravitational wave observatories let us now consider the correlated signals between two identical equal-arm Michelson interferometers to a stochastic gravitational wave background.  In this case we can form several different signals.  For example, the Michelson signals at vertices 1 and 3  can be written 
\begin{eqnarray}
s_{A}(t) &\equiv& \frac{1}{2} \left[\Delta \varphi_{12}(t-2L) + \Delta\varphi_{21}(t-L) - \Delta \varphi_{13}(t-2L) -\Delta \varphi_{31}(t-L)\right],\\
s_{C}(t) &\equiv& \frac{1}{2} \left[\Delta \varphi_{31}(t-2L) + \Delta\varphi_{13}(t-L) - \Delta \varphi_{32}(t-2L) - \Delta\varphi_{23}(t-L)\right].
\end{eqnarray}
The specific form of these Michelson signals have been chosen to ensure that their laser phase noise cancels \cite{Cornish:2001bb}. 
We are also interested in forming another signal, defined by 
\begin{equation}
s_{B}(t) \equiv s_{A}(t) + 2 s_{C}(t),
\end{equation}
where the `$B$' signal has been defined this way so that its total noise is uncorrelated with signal $A$ over the frequencies where the space-based interferometers are most sensitive (see Appendix~\ref{sec:noise}).  

The correlation between the signals (denoted by $X=A,B$ and $Y=A,B$) measured at each interferometer can be written as 
\begin{equation}
P^{XY}_{s}(f) = I(f) \mathcal{R}_I^{XY}(f) + V(f) \mathcal{R}_V^{XY}(f)
\end{equation}
where 
\begin{eqnarray}
\mathcal{R}_I^{XY}(f) &\equiv& \frac{1}{4\pi}\int d^2 \hat{n}  \left[F^X_{+}(\hat{n}, f) F^{*Y}_{+}(\hat{n}, f) + F^X_{\times}(\hat{n}, f) F^{*Y}_{\times}(\hat{n}, f)\right],\\
\mathcal{R}_V^{XY}(f) &\equiv& \frac{i}{4\pi} \int d^2 \hat{n}  \left[F^X_{+}(\hat{n}, f) F^{*Y}_{\times}(\hat{n}, f) - F^X_{\times}(\hat{n}, f) F^{*Y}_{+}(\hat{n}, f)\right],
\end{eqnarray}
and
\begin{equation}
F_{P}^X(\hat{n}, f) \equiv \mathcal{M}^{ab}_{X}(\hat{n},f) e^P_{ab}(\hat{n}).
\end{equation}
From the expressions for the response functions it is obvious that $\mathcal{R}_V^{AA} = \mathcal{R}_V^{BB} = 0$, $\mathcal{R}_I^{AB} =\mathcal{R}_I^{BA}$, and  $\mathcal{R}_V^{AB} = - \mathcal{R}_V^{BA}$. 
For the Michelson interferometer the transfer function $\mathcal{M}^{ab}_{X}(\hat{n},f)$ is given by \cite{Cornish:2001bb}
\begin{eqnarray}
\mathcal{M}^{ab}_{A}(\hat{n},f) \equiv   \frac{1}{2}e^{-2 \pi i f \hat{n} \cdot \vec x_1}  \left[(\hat \ell_{12}\otimes \hat \ell_{12}) \mathcal{F}_m(\hat \ell_{12} \cdot \hat{n},f) -(\hat \ell_{13} \otimes \hat \ell_{13}) \mathcal{F}_m(\hat \ell_{13} \cdot \hat{n},f)\right],\label{eq:transfer}\\
\mathcal{M}^{ab}_{B}(\hat{n},f) \equiv \mathcal{M}^{ab}_A(\hat{n},f) + e^{-2 \pi i f \hat{n} \cdot \vec x_3}\left[ (\hat \ell_{31} \otimes \hat \ell_{31}) \mathcal{F}_m(\hat \ell_{31} \cdot \hat{n},f) -(\hat \ell_{32} \otimes \hat\ell_{32}) \mathcal{F}_m(\hat \ell_{32} \cdot \hat{n},f) \right],
\end{eqnarray}
and
\begin{equation}
\mathcal{F}_m(\vec u \cdot \hat{n}, f) \equiv \frac{1}{2} \left[ {\rm sinc}\left( \frac{f(1-\vec u \cdot \hat{n})}{2 f_*} \right) \exp \left(- i \frac{f}{2 f_*} (3 + \vec u \cdot \hat{n}) \right) + {\rm sinc}\left(\frac{f (1+ \vec u \cdot \hat{n})}{2 f_*}\right) \exp \left(-i \frac{f}{2f_*} (1+\vec u \cdot \hat{n})\right) \right].
\end{equation}

We can now find signal combinations to form an estimator sensitive to the intensity and circular polarization of an ISGWB. The correlations are 
\begin{eqnarray}
\hat{\mathcal{C}}_I(f,f') &=& \left[ \tilde{s}_1^A(f)+\tilde{s}_1^B(f)\right] \left[\tilde{s}_2^{*A}(f')+\tilde{s}_2^{*B}(f')\right],\\
\hat{\mathcal{C}}_V(f,f') &=& \tilde{s}_1^A(f) \tilde{s}_2^{*B}(f')-\tilde{s}_1^{B}(f) \tilde{s}_2^{*A}(f'), 
\end{eqnarray}
which have expectation values 
\begin{eqnarray}
\langle \hat{\mathcal{C}}_I(f,f')\rangle &=& \frac{1}{2} I(f) \left[ \mathcal{R}_I^{AA}(f) +\mathcal{R}_I^{BB}(f) +2\mathcal{R}_I^{AB}(f) \right] \delta_T(f-f')\\ &\equiv& \frac{1}{2}S_{s,I}(f) \delta_T(f-f'), \nonumber \\
\langle \hat{\mathcal{C}}_V(f,f')\rangle &=& V(f)  \mathcal{R}_V^{AB}(f)\delta_T(f-f')\equiv \frac{1}{2}S_{s,V}(f) \delta_T(f-f').
\end{eqnarray}
As discussed in Appendix~\ref{sec:noise} the noise is uncorrelated between signal $A$ and $B$ so that the noise spectrum associated with each of these signals can be written 
\begin{eqnarray}
S_{n,I}(f) &=& \left[S_{n,A}(f) + S_{n,B}(f) \right]^2 \\
S_{n,V}(f) &=& S_{n,A}(f)S_{n,B}(f),
\end{eqnarray}
where we have 
\begin{eqnarray}
S_{n,A}(f) &=& 4\bigg(S_{n,s}(f) + 2 S_{n,a}(f)\left[1 + \cos^2(f/f_*)\right]\bigg), \\
S_{n,B}(f) &=&  \frac{3}{2} S_{n,A}(f),
\end{eqnarray}
and $S_{n,s}(f)$ and $S_{n,a}(f)$  is the shot-noise power spectrum and acceleration noise power spectrum, respectively. 

We calculate the sensitivity to circularly polarized gravitational waves for two planned space-based gravitational wave interferometers: the Laser Interferometer Space Antenna (LISA) \cite{eLISA}\footnote{We note that during a recent symposium in Zurich eLISA has been renamed LISA.} and BBO \cite{Crowder:2005nr}\footnote{Since both BBO and DECIGO are similar in design we only present noise curves for BBO. We also note that one should regard BBO/DECIGO as a straw-man design for the most sensitive gravitational wave detector in the $\sim$ Hz  frequency band using technology that is only slightly beyond the current state of the art.}.  The last stage of BBO calls for a six-mass configuration similar to what is shown in Fig.~\ref{fig:space-based}.  Current designs for LISA only include a three-mass equilateral configuration.  We include estimates for an `advanced' LISA with six masses in order to explore its potential sensitivity to circularly polarized gravitational waves. 

The parameters for LISA and BBO are shown in Table~\ref{tab:params}.  
\begin{table}[!tbh]
\begin{tabular}{l|c|c}
\hline
\hline
Parameter & LISA & BBO  \\
\hline
$L$ (m) & $10^9$ & $5 \times 10^7$   \\
\hline
$S_{n,s}(f)\ ({\rm Hz}^{-1})$ & $1.15\times10^{-40}$ & $8\times10^{-50}$  \\
\hline
$S_{n,a}(f)({\rm Hz}/f)^{-4}\ ({\rm Hz}^{-1})$ & $1.3 \times 10^{-50}(1+10^{-4} {\rm Hz}/f)$ & $2.3 \times 10^{-52} $  \\
\hline
$\Omega^I{\rm gw,min}$ & $5.0 \times 10^{-13}$ & $1.4 \times 10^{-17}$ \\
\hline
$\Omega^V{\rm gw,min}$ & $5.0 \times 10^{-13}$ & $1.4 \times 10^{-17}$ \\
\hline
\hline
\end{tabular}
\label{tab:space-based}
\caption{ 
Parameters for LISA from Ref.~\cite{eLISA}, BBO from Ref.~\cite{Crowder:2005nr}.} 
\label{tab:params}
\end{table}
The expression for the SNR for both the intensity and circular polarization allows us to determine the distance between the two observatories which maximizes the signal to noise to the circular polarization.  We have performed this calculation for LISA and BBO whose noise properties are described in Appendix \ref{sec:noise}.    
\begin{figure}[h!]
\begin{center}
\resizebox{!}{7cm}{\includegraphics{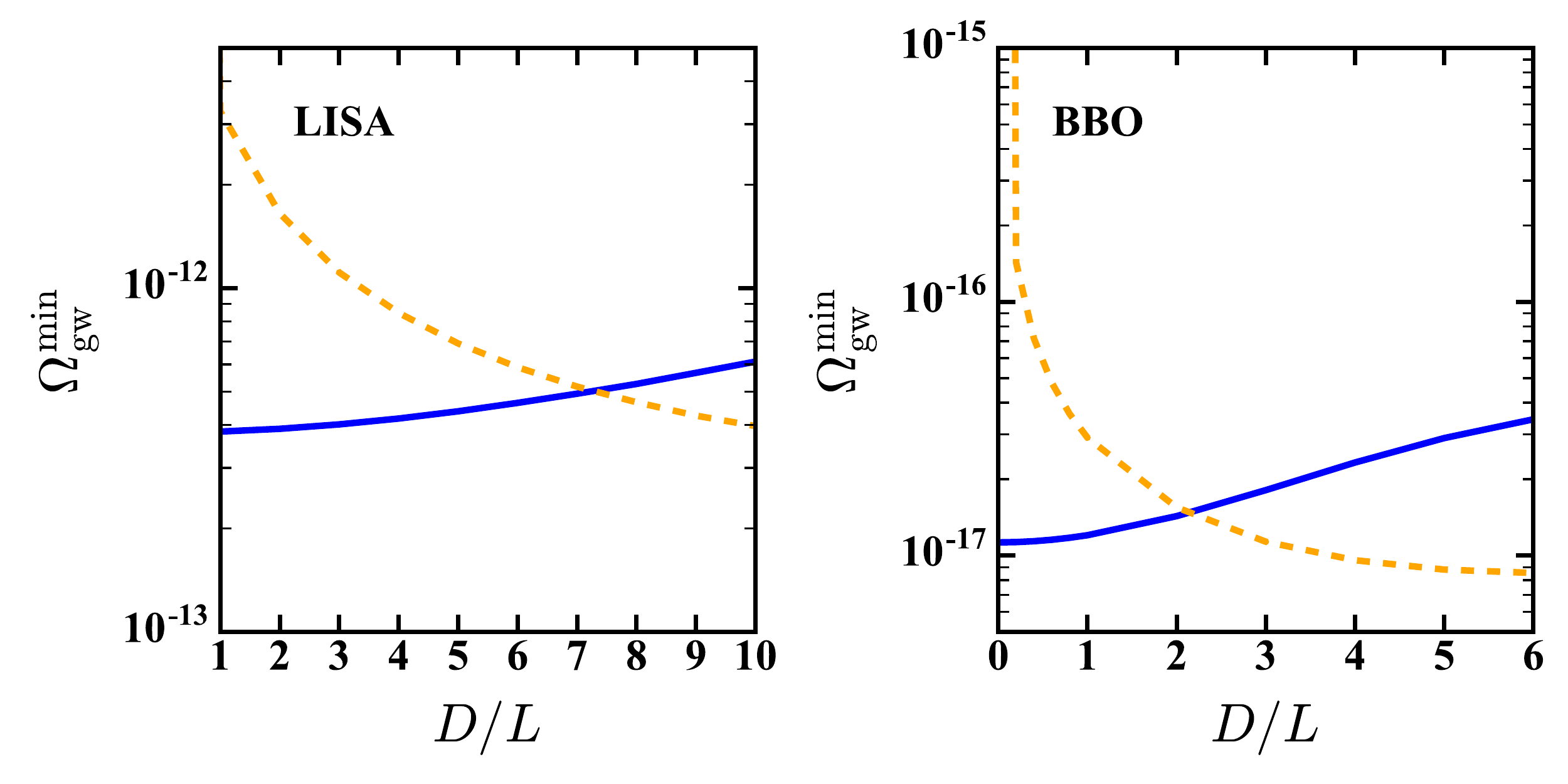}}
\caption{\emph{Left panel}: The minimum detectable $\Omega^{(I,V)}_{\rm gw}h^2$ as a function of the separation between the two observatories for LISA; sensitivity to the intensity $I$ is shown in solid blue, sensitivity to the circular polarization $V$ is shown in dashed orange.  \emph{Right panel}: The minimum detectable $\Omega^{(I,V)}_{\rm gw}h^2$ as a function of the separation between the two observatories for BBO; sensitivity to the intensity $I$ is shown in blue, sensitivity to the circular polarization $V$ is shown in orange. In both cases we have assumed a ten-year-long observation.}
\label{fig:minOmega}
\end{center}
\end{figure}
As shown in Fig.~\ref{fig:minOmega} we can see that both the intensity and circular polarization are detected at the same SNR for LISA if $D/L \simeq 7$ and for BBO if $D/L \simeq 2$.  This figure also shows that at these separations the sensitivity to the intensity, compared to $D=0$, is degraded by about 10\% in each case.  Since the arm-length of LISA has yet to be determined we also considered an arm-length of $L=2 \times 10^9$ m and $L=3 \times 10^9$ m.  We found that the overall sensitivity of these arm-lengths to the intensity and polarization is the same as when $L=10^9$ m but that they occur at smaller separations with $D/L = 3.75$ and  
 $D/L = 2.5$, respectively. 

We are now in a position to calculate the sensitivity curves for both observatories.  
\begin{figure}[h!]
\begin{center}
\resizebox{!}{7.2cm}{\includegraphics{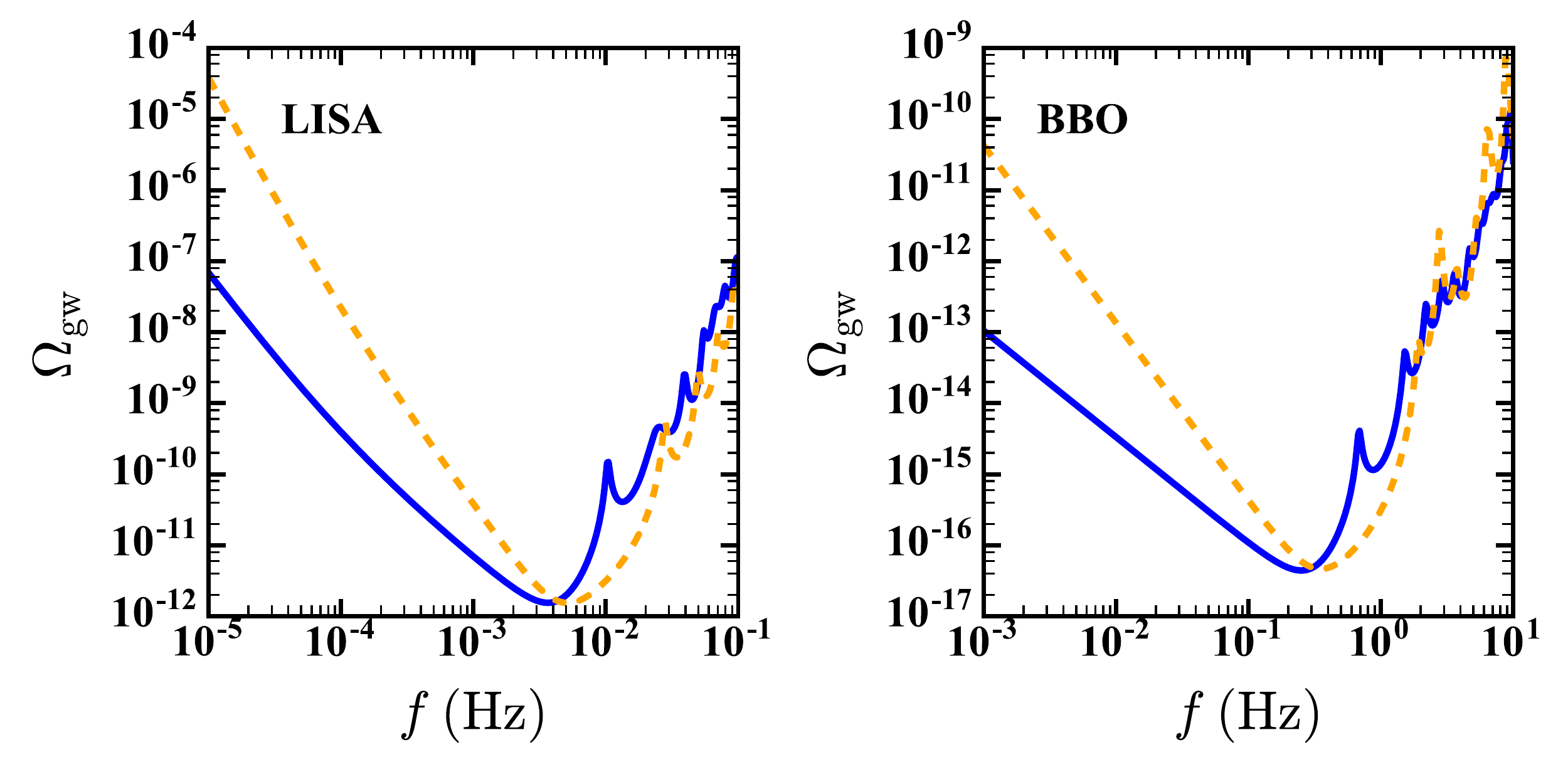}}
\caption{\emph{Left panel}: The sensitivity curve for $\Omega_{\rm gw}^{(I,V)}$ for LISA assuming a separation between the two observatories of $D/L = 7$, and  a 10-year-long observation.  The solid blue curve shows the sensitivity to the intensity and the dashed orange curve shows the sensitivity to the circularly polarized background.  Note that the sensitivity curve for $V$ has a smaller opening angle.  \emph{Right panel}: The same as in the left panel but for BBO and $D/L =2$.}
\label{fig:spacebased}
\end{center}
\end{figure}
Fig.~\ref{fig:spacebased} shows the frequency-dependent sensitivity curves for both LISA and BBO and Table \ref{tab:space-based}.  By construction these detectors are equally sensitive to the intensity and circular polarization, but there are important differences between the sensitivity curves. First, the curve for the 
circular polarization rises more sharply at low frequencies.  Expanding in small frequencies the intensity sensitivity curve rises as $f^{-5/2}$ 
whereas the circular polarization rises as $f^{-7/2}$.  We also find that the most sensitive frequency, $f^{\rm min}$, is shifted between the intensity and circular polarization with $f^{\rm min}_I = 3.6 \times 10^{-3}$ Hz, $f^{\rm min}_V = 5.1 \times 10^{-3}$ Hz for LISA and $f^{\rm min}_I = 0.25$ Hz, $f^{\rm min}_V = 0.35$ Hz for BBO. 

\section{Sensitivity of ground-based interferometers}
\label{sec:ground-based}

Ground-based interferometers monitor the relative phase at a single vertex.  A stochastic background would appear in the correlation between pairs of interferometers.  In the presence of a circularly polarized background the correlation between detector $i$ and $j$ takes the form 
\begin{equation}
\langle \tilde{s}_i(f) \tilde{s}^*_j(f')\rangle =\frac{1}{2}\left[ I(f) \mathcal{R}_I^{ij}(f) + V(f) \mathcal{R}_V^{ij}(f)\right] \delta_T(f-f').
\end{equation}
In order to extract the intensity and circular polarization information we need to consider the correlation between another pair of observatories, $k$ and $l$ (at least one of which needs to be different from  $i$ and $j$): 
\begin{eqnarray}
\hat{\mathcal{C}}^{ij,kl}_{(I,V)}(f,f') &=& \left(\frac{\tilde{s}_i(f) \tilde{s}^*_j(f')}{ \mathcal{R}^{ij}_{(V,I)}} -  \frac{\tilde{s}_k(f) \tilde{s}^*_l(f')}{ \mathcal{R}^{kl}_{(V,I)}}\right),
\label{eq:ground_based_est}
\end{eqnarray}
with the expectation values
\begin{eqnarray}
\langle \hat{\mathcal{C}}^{ij,kl}_{(I,V)}(f,f')\rangle &=& \frac{1}{2} \{I(f),V(f)\}  \left(\frac{\mathcal{R}^{ij}_{(I,V)}}{\mathcal{R}^{ij}_{(V,I)}} -\frac{\mathcal{R}^{kl}_{(I,V)}}{ \mathcal{R}^{kl}_{(V,I)}} \right)\delta_T(f-f').
\end{eqnarray}

We derive the optimal signal to noise for a network of 
ground-based observatories in Appendix \ref{sec:app_ground} and find that 
\begin{equation}
[({\rm SNR})_{(I,V)}]^2 =  T \sum_{i<j}\int  \frac{\{I(f),V(f)\} \left[\mathcal{R}_{(I,V)}^{ij}(f)\right]^2}{S_n^{(i)}(f) S_n^{(j)}(f)}df,
\end{equation}
where $S^{(i)}_n$ is the noise spectrum for each observatory.  In what follows we will assume that all observatories share the same noise spectrum given in Ref.~\cite{Adhikari:2013kya}. 
 We note that in order to separate the intensity from the 
level of circular polarization we need at least three well-separated observatories. 

Ground-based interferometers have noise spectra which restrict their sensitivity to frequencies $10\ {\rm Hz} \lesssim f \lesssim 10^3\ {\rm Hz}$.  The length of their arms is of order 1 km which corresponds to $f_* \simeq 10^6\ {\rm Hz}$. This means that we \emph{always} have $f/f_* \ll 1$ so that the transfer function in Eq.~(\ref{eq:transfer}) takes the simplified form \cite{Flanagan:1993ix}
\begin{equation}
\mathcal{T}^{ab} (\hat{n},f)  \simeq \frac{1}{2} \left(\hat{\ell}_{12}^a \hat{\ell}_{12}^b -\hat{\ell}_{13}^a \hat{\ell}_{13}^b\right).
\end{equation}
We note that with this approximation, the response of a ground-based observatory can be written analytically as shown in Refs.~\cite{Flanagan:1993ix,Seto:2008sr,Seto:2008sr2} and, for completeness, are reproduced in Appendix \ref{sec:app_ground}. 

The sensitivity of the world-wide network of ground-based gravitational wave observatories depends on the location of each observatory on the Earth as well as the relative orientation of their interferometer arms.  The location of each observatory is specified by its latitude and longitude and the orientation by the angle $\alpha$ which is measured counterclockwise from due east at each observatory (i.e.~the standard $\hat{\phi}$ in a spherical basis). 
\begin{table}[!bth]
\begin{tabular}{lccccc}
\hline\hline
 ~ & Lat & Long & $\alpha$ & $\Omega^I{\rm gw,min}$ &  $\Omega^V{\rm gw,min}$  \\
\hline
\ LIGO\ Hanford (H) & $46.45$ & $-119.41$ & $171$ &--& -- \\
\ LIGO\ Livingston (L) & $30.56$ & $-90.77$ & $242$ &--&-- \\
\ Virgo (V) & $43.63$ & $10.5$ & $115.6$ &  $1.3 \times 10^{-10}$ & $4.6 \times 10^{-10}$ \\
\ LIGO\ India (I) & $10.02$ & $77.76$ & $58.2$ &--&-- \\
\ KAGRA (K) & $36.42$ & $137.3$ & $75$ &$1.1 \times 10^{-10} $& $2.0 \times 10^{-10}$ \\
\hline\hline
\end{tabular}
\caption{Positions $({\rm Lat,Long})$ and orientation angles $\alpha$ (all in degrees) of
the ground-based detectors considered in this paper. The minimum detectable $\Omega_{\rm gw}$ includes all previously listed observatories.  When all five observatories are correlated the sensitivity to the intensity of the ISGWB is improved by a factor of approximately two whereas the sensitivity to the 
 level of circular polarization is improved by about a factor of three.}
\label{tab:location_of_detectors}
\end{table}
In order to disentangle the intensity and circular polarization we need to consider the correlation between two other interferometers.  In this case we choose to use LIGO-Hanford (H), LIGO-Livingston (L), LIGO-India (I), Virgo (V), and KAGRA (K) (previously known as the Large Scale Cryogenic Gravitational Wave Telescope \cite{2010CQGra..27h4004K}).  Of these observatories LIGO-India's location and orientation has yet to be determined.  For LIGO-India we take the location and orientation determined in Ref.~\cite{2013CQGra..30o5004R} to optimize the polarization reconstruction and effective angular resolution of a multi-observatory detection of a periodic source. 
\begin{figure}[h!]
\begin{center}
\resizebox{!}{7.5cm}{\includegraphics{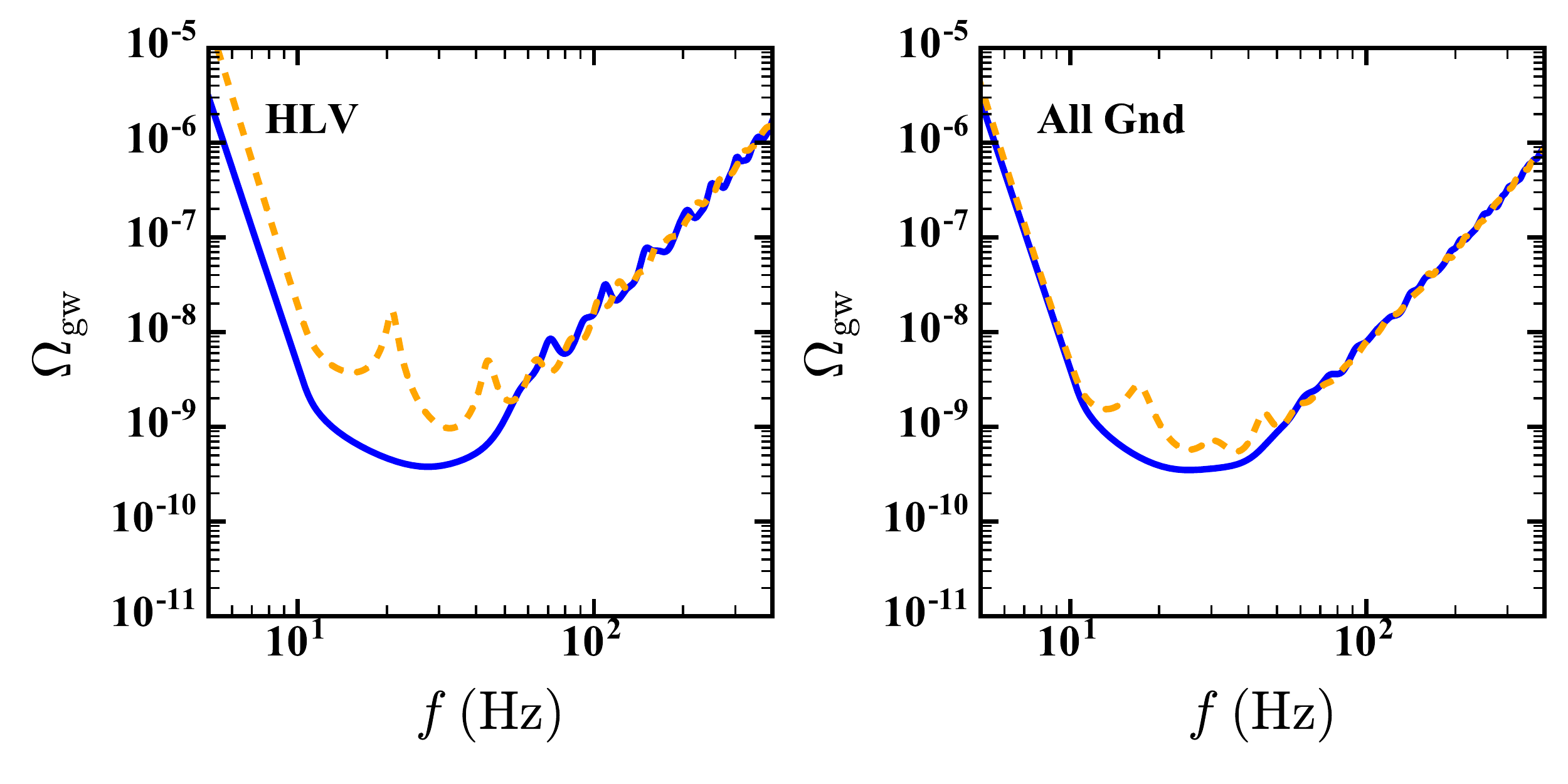}}
\caption{\emph{Left panel}: The sensitivity of currently built ground-based observatories (LIGO Hanford and Livingston, Virgo).  The solid blue and dashed orange curves show the sensitivity to the intensity and circular polarization. \emph{Right panel}: The sensitivity curve for the five current and planned ground-based observatories listed in Table \ref{tab:location_of_detectors} and a 10-year-long observation.}
\label{fig:minOmega_ground}
\end{center}
\end{figure}

We show the sensitivity to both the intensity and circular polarization of a stochastic gravitational wave background in Fig.~\ref{fig:minOmega_ground}. First we consider correlations between the signal measured by the LIGO Hanford, LIGO Livingston, and Virgo observatories, shown in the left panel of Fig.~\ref{fig:minOmega_ground}.  With this limited set of observatories there is a significant difference between the sensitivities to the intensity and the circular polarization, with the sensitivity to the intensity about four times greater than to the circular polarization.  When we include LIGO India and KAGRA this difference is reduced to about a factor of two.  The curves in Fig.~\ref{fig:minOmega_ground} also show that the sensitivity to the circular polarization does not have a smooth minimum, but instead varies significantly with frequency.  This needs to be taken into account when assessing the ability for a ground-based network to detect a frequency dependent circular polarization. 

\subsection{The Einstein Telescope}

The Einstein Telescope (ET) is a proposed next-generation ground-based gravitational wave observatory. It is currently planned to consist of six Michelson interferometers each with an opening angle of $60^\circ$ oriented relative to each other to form an equilateral triangle.  Three of the interferometers are designed to optimize sensitivity to `high' frequency (HF) gravitational waves ($f \sim 10 -10^4$ Hz) and three to optimize `low' frequencies ($f \sim 1-250$ Hz).  The arm-length of these interferometers will be 10 km, compared to the 3-4 km arm-lengths of current observatories.  Furthermore, the ET is planned to be built underground in order to better isolate it from seismic activity.  Based on measurements of seismic activity, the ET may be built at one of three sites in Europe (Gy{\"o}ngy{\"o}soroszi mine, Hungary; LSC, Canfranc, Spain; Sos Enattos mine, Sardinia, Italy).  
\begin{table}[!bth]
\begin{tabular}{lcccc}
\hline\hline
 ~ & Lat & Long & $\Omega^I{\rm gw,min}$ &  $\Omega^V{\rm gw,min}$\\
\hline
Gy{\"o}ngy{\"o}soroszi mine & $47.78$ & $19.93$ &$8.63 \times 10^{-14}$ &$ 2.0 \times 10^{-11}$ \\
Canfranc, Spain & 42.71 & 0.52 & $8.63 \times 10^{-14}$ & $1.78 \times 10^{-11}$ \\
Sos Enattos mine & 40.47 & 9.483 &$8.63 \times 10^{-14}$ & $1.89 \times 10^{-11}$  \\
\hline\hline
\label{tab:ET}
\end{tabular}
\caption{Positions $({\rm Lat,Long})$ and orientation angles $\alpha$ (all in degrees) of
 the ground-based detectors considered in this paper.}
\label{tab:location_of_detectors}
\end{table}
Since the ET forms an equilateral triangle, we take the signal it will measure as being produced at three vertices each with a noise spectral density given by the ET-D configuration described in Ref.~\cite{Hild:2010id}.  
\begin{figure}[h!]
\begin{center}
\resizebox{!}{7.5cm}{\includegraphics{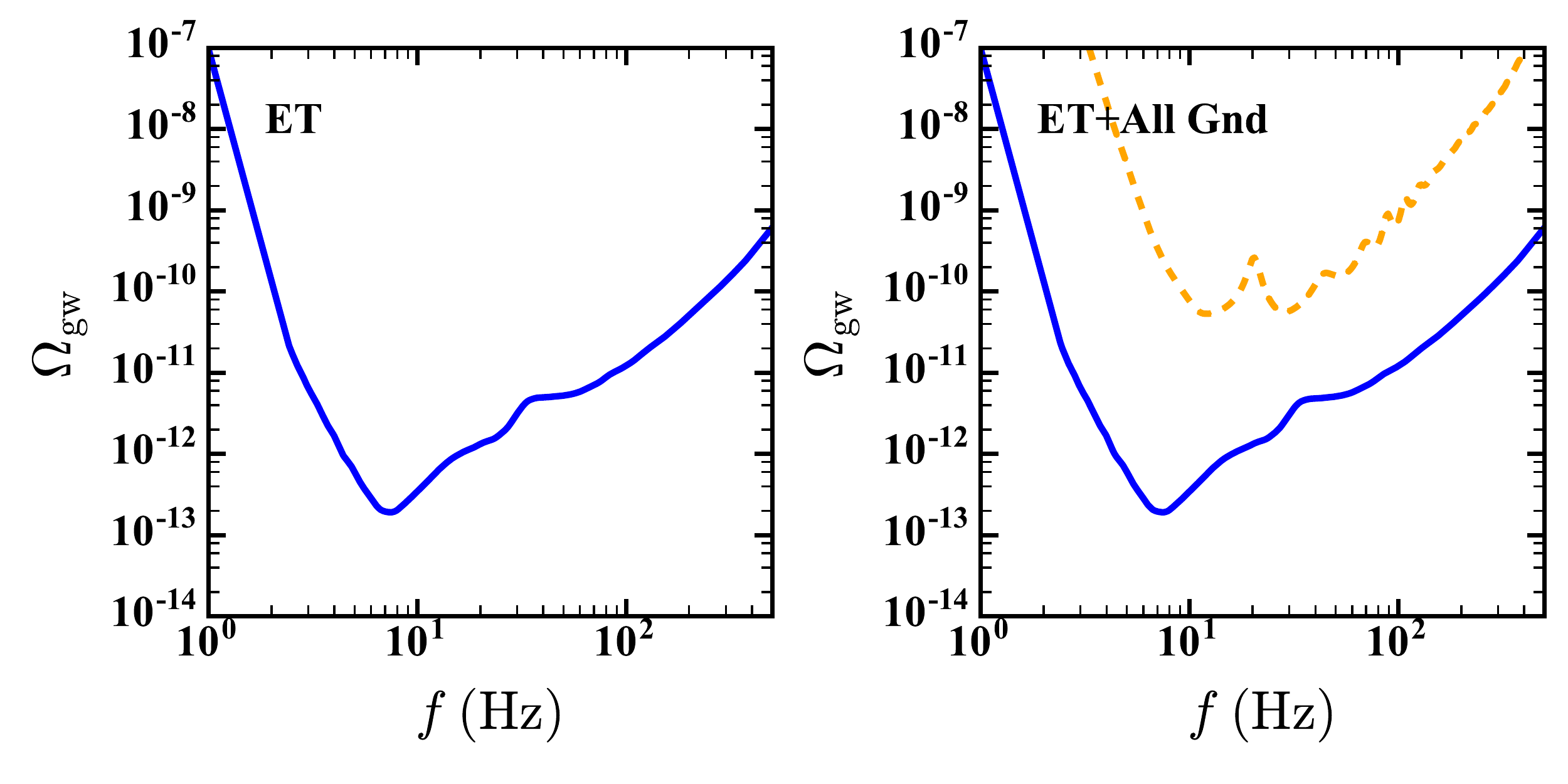}}
\caption{\emph{Left panel}: The sensitivity of the Einstein Telescope (ET).  Since the ET consists of phase measurements at the vertices of an equilateral triangle, it is not intrinsically sensitive to the level of circular polarization in the ISGWB. \emph{Right panel}: When we correlate the ET signal with other current and planned ground-based gravitational wave observatories the network is sensitive to the circular polarization, shown in the dashed-orange curve.}
\label{fig:minOmega_groundET}
\end{center}
\end{figure}

The correlated signal between the three vertices of the ET is sensitive to the intensity of the ISGWB but, because the vertices are coplanar, is insensitive to circular polarization. However, when the signals are correlated with the global ground-based gravitational wave observatory network, the addition of the ET greatly improves the network's sensitivity to both the intensity and circular polarization of the ISGWB, as shown in Table \ref{tab:ET}. 

We investigated whether one of the possible sites for the ET yielded a network of ground-based observatories with significantly improved sensitivity over the other two.  We also investigated whether a particular orientation for the ET maximized the network's sensitivity.  Although locating the ET at Canfrac, Spain yields a marginally more sensitive network to circularly polarized gravitational waves, the improvement is minimal.  We also investigated whether particular orientations of ET would yield a more sensitive network and found that changing the orientation has a negligible effect on the sensitivity to the intensity and can change the sensitivity to the level of circular polarization by at most 10\%. 

It is interesting to note that the three sites considered for the ET, along with the current and planned sites for the other gravitational wave observatories, is highly concentrated in the northern hemisphere.  We explored the possibility of building an ET-like observatory in the southern hemisphere (near Pretoria, South Africa with latitude $25.7^{\circ}$ S and longitude $28.2^\circ$ E. ) and found that the overall sensitivity to the intensity is  unchanged with $\Omega^I_{\rm gw, min} = 8.63\times10^{-14}$ but the level of circular polarization is improved by more than a factor of two, $\Omega^V_{\rm gw,min} = 7.92\times10^{-12}$, as compared to the values in Table~\ref{tab:location_of_detectors}.  

\section{CMB sensitivity} 

At the largest scales, measurements of the CMB provide us with a tool that has the potential to detect gravitational waves at frequencies $f \simeq 10^{-18}\ {\rm Hz}$.  The presence of gravitational waves on these scales induces correlated fluctuations in both the intensity and polarization of the CMB \cite{Kamionkowski:1996zd,Seljak:1996gy}.  Expanding the intensity and polarization measurements in the appropriate spin-weighted multipole moments, we can write the gravitational-wave induced correlations as integrals over $I(k)$ and $V(k)$:
\begin{eqnarray}
C_\ell^{XX'=TT,EE,BB,TE} &=& (4 \pi)^2\int k^2 dk \, I(k) \Delta^{{\rm GW}, X}_\ell(k)  \Delta^{{\rm GW}, X'}_\ell(k),\\
C_\ell^{XX'=TB,EB} &=& (4 \pi)^2\int k^2 dk \, V(k) \Delta^{{\rm GW}, X}_\ell(k)  \Delta^{{\rm GW}, X'}_\ell(k),
\end{eqnarray}
where $\Delta^{{\rm GW}, X}_\ell(k)$ are the transfer functions which encode the physics of photon transport from the surface of last scattering to today. The noise   
at each multipole can be written 
\begin{equation}
N^{X_1X_2,X_3 X_4}_\ell = \frac{1}{2 \ell +1} \left( \tilde C_{\ell}^{X_1 X_3} \tilde C_{\ell}^{X_2 X_4} + \tilde C_{\ell}^{X_1 X_4} \tilde C_{\ell}^{X_2 X_3}\right),
\end{equation}
with
\begin{equation}
\tilde C_{\ell}^{X X'}  \equiv C_\ell^{XX',s} +  \delta_{XX'}\frac{4 \pi \sigma_X^2}{N_{\rm pix}} e^{\ell^2 \sigma_b^2}, 
\end{equation}
where $C_\ell^{XX',s}$ is the scalar (i.e., non gravitational-wave) contribution to the power spectrum, $\sigma_b \equiv \theta_{FWHM}/\sqrt{8 \ln 2}$ and we have assumed that the cross-correlated noise vanishes, $\sigma_X$ is the pixel noise, and $N_{\rm pix} = 4 \pi \theta_{\rm FWHM}^{-2}$ is the number of pixels. Under the null hypothesis a non-zero $C_\ell^{BB}$ is produced through weak lensing and $C_\ell^{TB,s} = C_{\ell}^{EB,s} = 0$. 
\begin{figure}[h!]
\begin{center}
\resizebox{!}{7.5cm}{\includegraphics{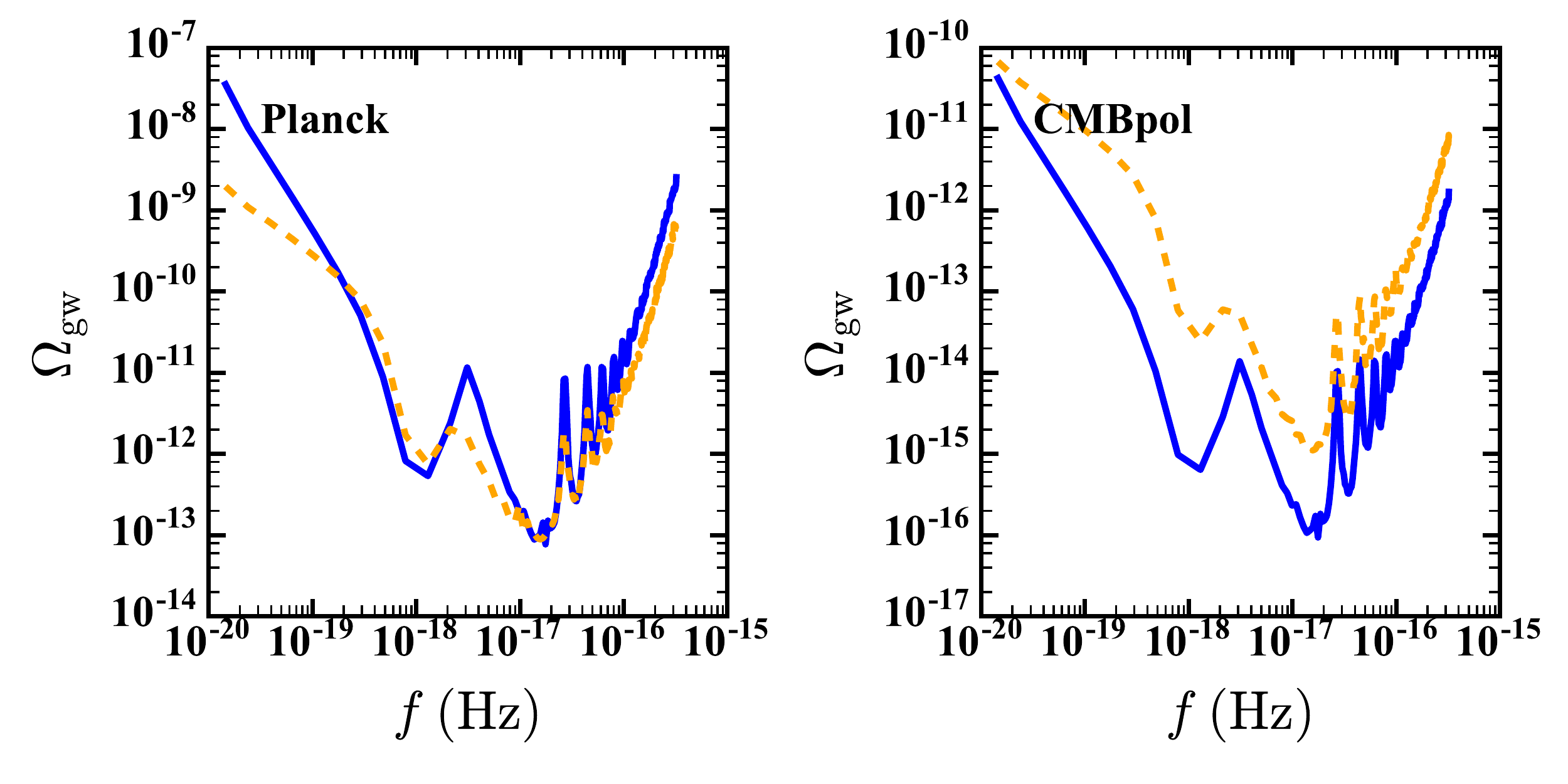}}
\caption{The sensitivity of CMB observations to the intensity (solid blue) and circular polarization (dashed orange) for both Planck (left panel) and CMBpol (right panel).  Our noise model and parameters for these two instruments are specified in Appendix \ref{sec:CMB}.}
\label{fig:CMB}
\end{center}
\end{figure}

From the expression for the noise covariance under the null hypothesis we have 
\begin{eqnarray}
N^{TT,BB}_{\ell} &=& 0,\\
N^{TB,EB}_{\ell} &=&  \frac{\tilde{C}_{\ell}^{T E,s} \tilde C_{\ell}^{BB}}{2 \ell +1},\\
N^{EB,EB}_{\ell} &=& \frac{ \tilde C_{\ell}^{E E,s} \tilde C_{\ell}^{BB}}{2 \ell +1},\\
N^{TB,TB}_{\ell} &=& \frac{ \tilde C_{\ell}^{T T} \tilde C_{\ell}^{BB}}{2 \ell +1}.
\end{eqnarray}
The SNR for the ISGWB intensity from CMB experiments is dominated by the $TT$ and $BB$ measurements so that 
\begin{eqnarray}
({\rm SNR})_I^2 &=& \sum_\ell \frac{[C_{\ell}^{TT,GW}]^2}{N_\ell^{TT,TT}}+\frac{[C_{\ell}^{BB,GW}]^2}{N_\ell^{BB,BB}},\label{eq:CMBI}\\
({\rm SNR})_V^2 &=& \sum_\ell \frac{[C_{\ell}^{TB}]^2N^{EB,EB}_{\ell} - 2 C_\ell^{EB} C_\ell^{TB} N^{TB,EB}_{\ell} + [C_\ell^{EB}]^2N_\ell^{TB,TB} }{N_\ell^{EB,EB} N_\ell^{TB,TB}-[N_\ell^{TB,EB}]^2}.\label{eq:CMBV}
\end{eqnarray}

In order to estimate the sensitivity curve of the CMB to the intensity and circular polarization of the ISGWB we note that the transfer function peaks at a wavenumber $\ell = k_{\ell} \tau_0$ where $\tau_0$ is the conformal time today. This allows us to 
write the power spectra as a function of wavenumber $k_\ell$:
\begin{eqnarray}
C_\ell^{XX'} &\simeq& (4 \pi)^2 \{I(k_\ell),V(k_\ell)\} \tau_0^{-1} k_\ell^2 \Delta_\ell^{X}(k_\ell)  \Delta_\ell^{X'}(k_\ell).
\end{eqnarray}
Therefore the SNR can be written as the sum of the square of each of the scale-dependent SNRs:
\begin{equation}
{\rm SNR}_{(I,V)}^2 \simeq \sum_{\ell} {\rm SNR}^2_{(I,V)}(k_\ell).
\end{equation}
Using the same approach described in the previous sections this allows us to calculate the frequency-dependent sensitivity of CMB observations (after noting that $k = 2 \pi f/c$) to both the intensity and the circular polarization of a ISGWB as detailed in Appendix \ref{sec:CMB} and shown in Fig.~\ref{fig:CMB}.

\begin{table}[!bth]
\begin{tabular}{lccccc}
\hline\hline
 ~ & $\Theta_{\rm FWHM}$ & NET ($\mu{\rm K} \sqrt{s}$) & $T_{\rm obs}$ (years)& $\Omega^I{\rm gw,min}$ &  $\Omega^V{\rm gw,min}$\\
\hline
Planck & $7$ & $62$ & 1.2 &$1.53\times 10^{-14}$ &$ 1.76   \times 10^{-14}$ \\
CMBpol & 5  & 2.8 & 4&$2.13 \times 10^{-17}$ & $2.19 \times 10^{-16}$ \\
\hline\hline
\end{tabular}
\caption{Noise parameters for the Planck satellite \cite{planck} and CMBPol \cite{Bock:2008ww}.  Since each observation is of the full sky we take $f_{\rm sky} = 0.7$ to account for the subtraction of the galaxy.}
\label{tab:cmb_params}
\end{table}

The oscillations in these sensitivity curves follow the acoustic oscillations in the spectra.  In particular, the significant increase in sensitivity around $f \simeq 10^{-18}\ {\rm Hz}$ corresponds to the reionization bump at $\ell \simeq 4$ and the second dip corresponds to the horizon at decoupling at $\ell \simeq 100$.  It is also interesting to note that Planck is equally (in)sensitive to the intensity and circular polarization of the ISGWB whereas CMBpol is significantly more sensitive to the intensity. 

\section{Conclusions}

\begin{figure}[h!]
\begin{center}
\resizebox{!}{7.5cm}{\includegraphics{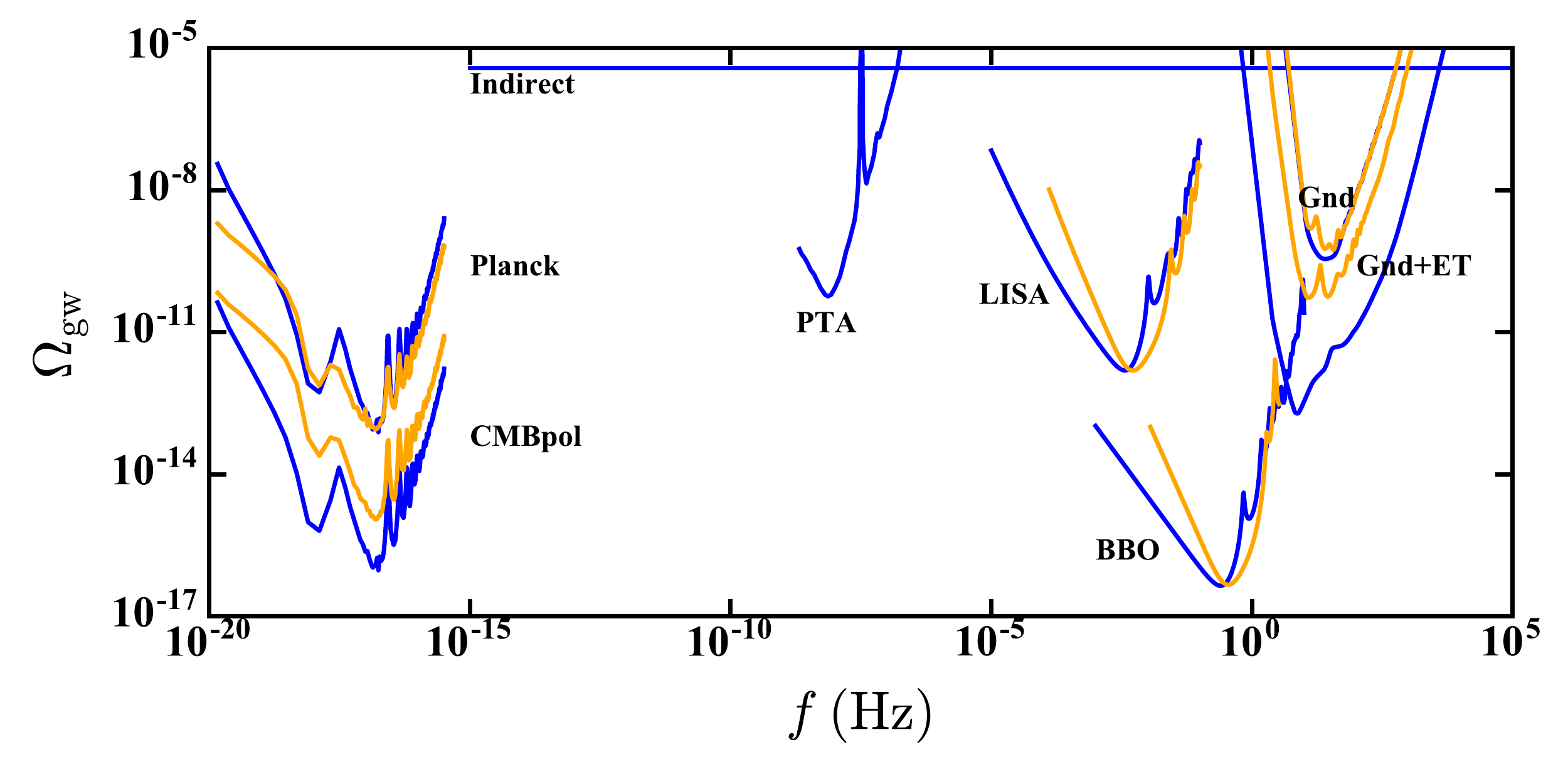}}
\caption{A collection of all of the sensitivity curves calculated in this paper with solid blue giving the sensitivity to the intensity and dashed orange to the level of circular polarization.  We have also included the sensitivity to the ISGWB from pulsar timing array from Ref.~\cite{Ellis:2012zv} and indirect limits coming from CMB measurements of the radiative energy density of the universe \cite{Smith:2006nka,Pagano:2015hma}.}
\label{fig:ALL}
\end{center}
\end{figure}

As shown in this paper, most of the common techniques used to detect the ISGWB will be sensitive to both the intensity and level of circular polarization.  We have summarized the sensitivity curves calculated in this paper in Fig.~\ref{fig:ALL}: the solid blue curves show the sensitivity to the intensity and the dashed orange curves show the sensitivity to the level of circular polarization.  

Space-based detectors will be sensitive to both the intensity and circular polarization as long as they utilize more than three inertial masses.  We have considered the case where these detectors operate as a constellation of two equilateral triangles.  The two triangles must be separated by some distance, and there is a distance at which the overall sensitivity to both the intensity and circular polarization are equal, in agreement with Ref.~\cite{Seto:2006dz}.  In addition to this we found that this optimal distance has a strong dependence on the specifications of the observatory -- for LISA we found that the optimal distance $D \simeq 7L$ whereas for BBO $D \simeq 2L$. 

Ground-based detectors are sensitive to both the intensity and circular polarization as long as we correlate the signal from at least three widely separated sites.  This means that the current collection of ground-based detectors (LIGO Hanford and LIGO Livingston) are not capable of separating out these two signals.  However, with VIRGO and KAGRA soon to turn on, the ground-based network will become sensitive to both signals.  We find that this total network sensitivity is greatly enhanced if we include the Einstein Telescope.  Since the intrinsic sensitivity of the Einstein Telescope to the intensity is significantly better than current gravitational wave observatories, it has a disproportionate effect on the overall sensitivity to the intensity.  However, it also significantly improves the network's sensitivity to the level of circular polarization.  We also found that if we were to locate the Einstein Telescope in the southern hemisphere the improvement in the total sensitivity to the level of circular polarization further improves by another factor of two. 

Observations of the temperature and polarization of the CMB are sensitive to both the intensity and circular polarization of the ISGWB.  The correlation between the CMB temperature and the $E$ and $B$ mode polarization can isolate the effects of the ISGWB intensity from those of the circular polarization.  In agreement with Ref.~\cite{Gluscevic:2010vv} we find that the Planck satellite is equally (in)sensitive to the intensity and circular polarization of the ISGWB, but that a future CMB satellite dedicated to measuring the CMB polarization -- CMBpol -- will improve the sensitivity by three orders of magnitude for the intensity of the ISGWB and two orders of magnitude for the circular polarization. 

As opposed to reporting the sensitivity as a single number, the calculation of sensitivity curves gives a quantitative accounting of the frequency coverage by these various observatories.  Looking at the combination of all of the observatories considered in this paper in Fig.~\ref{fig:ALL}, it is interesting to note the absence of any detector operating at frequencies between $10^{-15}\ {\rm Hz} \lesssim f\lesssim 10^{-3}\ {\rm Hz}$ which will be sensitive to the level of circular polarization. This 18 orders of magnitude is a wide swath of frequency space inside of which we do not have any known technique to detect the circular polarization of the gravitational wave background.  This sensitivity desert calls out for new and creative ideas on how to detect the level of circular polarization in an ISGWB. 

\begin{acknowledgments}
We thank Naoki Seto and Scott Hughes for useful discussions.  TLS thanks Matt Zucker for useful discussions about the SLERP algorithm. Work at Dartmouth was supported in part by DOE grant SC0010386. TLS worked on the initial stages of this project at the Aspen Center for Physics, which is supported by National Science Foundation grant PHY-1066293.
\end{acknowledgments}

\begin{appendix}

\section{Noise in a space-based laser interferometer \label{sec:noise}} 

The interferometer signal is built out of phase measurements made at each detector.  These measurements take the difference between the incoming light and the local light signal.  A gravitational wave interferometer will have three (major) sources of noise: the laser phase noise, $C(t)$, shot noise, $n^s(t)$, and acceleration noise, $\vec n^a(t)$.  If we denote the phase measurement made by detector $j$ with a laser sent by detector $i$ by $N_{ij}$ then \cite{Cornish:2001bb}
\begin{equation}
N_{ij}(t) = C_i(t-L_{ij}) - C_j(t) + n_{ij}^s(t) - \hat{\ell}_{ij} \cdot \left[\vec n_{ij}^a(t) - \vec n^a_{ji}(t-L_{ij})\right].
\end{equation}
For the equal-arm Michelson interferometer the laser phase noise cancels and we have the autocorrelation of the detector noise at vertices 1 and 2 (signals $A$ and $C$, respectively)
\begin{eqnarray}
\langle \tilde{N}_A(f) \tilde{N}^*_A(f')\rangle &=&  4\bigg(S_s(f) + 2 S_a(f)\left[1 + \cos^2(f/f_*)\right]\bigg) \delta(f-f') \equiv \frac{1}{2} P_N(f) \delta(f-f'),\\
\langle \tilde{N}_C(f) \tilde{N}^*_C(f')\rangle &=& \frac{1}{2} P_N(f) \delta(f-f').
\end{eqnarray}
There is a non-zero cross-correlation between $\tilde{N}_A$ and $\tilde{N}_C$ because of the common `arm' between vertices 1 and 3.  A full calculation of this cross correlation using the expressions in Ref.~\cite{Cornish:2001bb} yields 
\begin{eqnarray}
\langle \tilde{N}_A(f) \tilde{N}^*_C(f')\rangle &=& -2 \big(4 S_a(f) + S_s(f)\big) \cos(2 f/f_*),\\
&\simeq& -\frac{1}{4} P_N(f) \delta(f-f'),
\end{eqnarray}
where the approximate equality is accurate when $\cos(2 f/f_*)\simeq 1$.  Since this is true at the most sensitive frequencies of both LISA and BBO it is a good approximation when calculating the optimal SNR for these interferometers. 
Now for the $B$ signal we have 
$N_B(t) = N_A(t) + 2N_C(t)$
so that 
\begin{eqnarray}
\langle \tilde{N}_B(f) \tilde{N}^*_B(f')\rangle &=&  \langle \tilde{N}_A(f) \tilde{N}^*_A(f')\rangle +4 \langle \tilde{N}_C(f) \tilde{N}^*_C(f')\rangle+4\langle \tilde{N}_A(f) \tilde{N}^*_C(f')\rangle,\\
&=& \frac{3}{2} P_N(f) \delta(f-f') ,\\
\langle \tilde{N}_A(f) \tilde{N}^*_B(f')\rangle &=& \langle \tilde{N}_A(f)[ \tilde{N}^*_A(f')+2N^*_C(f')]\rangle = 0.
\end{eqnarray}

\section{Ground-based interferometer-network response and noise \label{sec:app_ground}}

As discussed in Sec.~\ref{sec:ground-based}, for each pair of pairs we can form an estimator of the intensity and circular polarization 
\begin{equation}
    \hat{\mathcal{C}}_{(I,V)}^{ij,kl}(f,f') = \frac{\tilde{s}_i(f) \tilde{s}^*_j(f')}{\mathcal{R}_{(V,I)}^{ij}(f)} - \frac{\tilde{s}_k(f) \tilde{s}^*_l(f')}{\mathcal{R}_{(V,I)}^{kl}(f)}.
\end{equation}
We can then form the frequency-integrated estimator 
\begin{equation}
    \hat{\mathcal{C}}_{(I,V)}^{ij,kl} = \int df df' W^{(I,V)}_{ij,kl}(f,f')\hat{\mathcal{C}}_{(I,V)}^{ij,kl}(f,f'),
\end{equation}
which has the expectation value 
\begin{equation}
\big\langle   \hat{\mathcal{C}}_{(I,V)}^{ij,kl} \big\rangle  = \frac{1}{2}\int df df' \{I(f),V(f)\} W^{(I,V)}_{ij,kl}(f,f')\left(\frac{\mathcal{R}^{ij}_{(I,V)}(f)}{\mathcal{R}^{ij}_{(V,I)}(f)} -\frac{\mathcal{R}^{kl}_{(I,V)}(f)}{ \mathcal{R}^{kl}_{(V,I)}(f)}\right) \delta_T(f-f').
\end{equation}
We note that even though the expectation value $\langle \hat{\mathcal{C}}^{ij,kl}_{(I,V)}(f,f')\rangle$ is not positive definite, the optimal estimator derived in Sec.~\ref{sec:opSNR} weights these terms to ensure that they always contribute positively to the overall signal to noise. 

Any ground-based network with more than two observatories will have more than one pair of pairs.\footnote{For example the LIGO Hanford (H) and Livingston (L) sites along with VIRO (V) provide three pairs of pairs: HL-HV, HV-LV, HL-LV.} In this case we can improve the SNR by combining all possible correlations:
\begin{equation}
\hat{\mathcal{C}}_{(I,V)} \equiv \sum_{ij,kl}\int df df' W^{(I,V)}_{ij,kl}(f,f')\hat{\mathcal{C}}_{(I,V)}^{ij,kl}(f,f') ,
\end{equation}
where the sum is over unique pairs of pairs without regard to order.  

By collecting the terms involving each estimator pair we can write this in the form 
\begin{equation}
\hat{\mathcal{C}}_{(I,V)} = \frac{1}{2} \int_{-\infty}^{\infty} df df' W_{(I,V)}^{ij}(f,f') \tilde{s}_i(f) \tilde{s}_j^*(f'),
\end{equation}
where $W_{(I,V)}^{ij}(f,f')$ has the properties discussed in Sec.~\ref{sec:opSNR} and is a linear combination of the weights $W_{ij,kl}(f,f')$ that involve detectors $ij$.
For example with the three detectors at LIGO Hanford (H), LIGO Livingston (L), and Virgo (V):
\begin{eqnarray}
W_{(I,V)}^{HL}&=& \frac{1}{\mathcal{R}_{(V,I)}^{HL}} \left(W_{(I,V)}^{HL,HV} + W_{(I,V)}^{HL,LV}\right),\\
W_{(I,V)}^{HV}&=& \frac{1}{\mathcal{R}_{(V,I)}^{HV}} \left(W_{(I,V)}^{HV,LV} - W_{(I,V)}^{HL,HV}\right),\\
W_{(I,V)}^{LV}&=& -\frac{1}{\mathcal{R}_{(V,I)}^{LV}} \left(W_{(I,V)}^{HL,LV} +W_{(I,V)}^{HV,LV}\right).
\end{eqnarray}
Given $N_O$ observatories, for each of the $N_P=1/2N_O(N_O-1)$ pairs there is a weight $W_{ij}$ that we construct using $2(N_O-2)$ of the 
$N_{PP}=N_P(N_O-2)$ pairs of pairs $W_{ij,kl}$.
With this it is then straightforward to show that the expectation value of this estimator is 
\begin{equation}
\langle \hat{\mathcal{C}}_{(I,V)} \rangle =  \frac{1}{4} \int_{-\infty}^{\infty} df df' W^{(I,V)}_{ij}(f,f') \mathcal{R}^{ij}_{(I,V)}(f) \delta_T(f-f').
\end{equation}
As shown in Sec.~\ref{sec:opSNR} the optimal SNR for this estimator is then given by 
\begin{equation}
{\rm SNR}_{(I,V)} = \left[T \sum_{i<j} \int_{-\infty}^\infty \{I(f),V(f)\}\frac{\left[\mathcal{R}^{ij}_{(I,V)}(f)\right]^2}{S_n^{(i)}(f) S_n^{(j)}(f)}\right]^{1/2}.
\end{equation}

As shown in Refs.~\cite{Flanagan:1993ix,Seto:2008sr} the ground-based response functions can be written down analytically:
\begin{eqnarray}
\mathcal{R}_I^{ij} &=& \frac{4}{5} \left[\Theta_1(y,\beta) \cos(4 \delta) + \Theta_2(y,\beta) \cos(4\Delta)\right], \\
\mathcal{R}_V^{ij} &=& \frac{4}{5}   \Theta_3(y,\beta) \sin(4\Delta), 
\end{eqnarray}
where $\delta \equiv \frac{\sigma_1 - \sigma_2}{2}$, $\Delta \equiv \frac{\sigma_1+\sigma_2}{2}$,
\begin{eqnarray}
\Theta_1(y, \beta) &\equiv& \cos^4 \frac{\beta}{2} \left( j_0(y) + \frac{5}{7} j_2(y) + \frac{3}{112} j_4(y) \right), \\
\Theta_2(y, \beta) &\equiv& \left(- \frac{3}{8} j_0(y) + \frac{45}{56} j_2(y) - \frac{169}{896} j_4(y)\right) + \left(\frac{1}{2} j_0(y) - \frac{5}{7} j_2(y) - \frac{27}{224} j_4(y) \right) \cos \beta\nonumber \\ &+& \left(-\frac{1}{8} j_0(y) - \frac{5}{56} j_2(y) - \frac{3}{896} j_4(y) \right) \cos 2 \beta, \\
\Theta_3(y,\beta) &\equiv& - \sin \frac{\beta}{2} \left[-j_1(y) + \frac{7}{8} j_3(y) + \left(j_1(y) + \frac{3}{8} j_3(y)\right)\cos \beta \right],
\end{eqnarray}
$j_n(y)$ is the ${\rm n}^{\rm th}$ spherical Bessel function, $y \equiv 2 F  \sin \beta/2$, and $F \equiv f/f_*$ with $f_* \equiv c/(2\pi R_{\rm E})$. 

Since we are only interested in the isotropic background the relative position of any two observatories on the surface of the Earth is characterized by three angles: $\beta$ is the angular separation between the two corner detectors measured from the center of the Earth and $\sigma_{a,b}$ which indicates the angular orientation of the bisector of the interferometer as measured counterclockwise relative to the great circle that connects the two observatories. The distance between the two observatories is $D = 2 R_E \sin \beta/2$.  We can establish these angles by imagining the two observatories as starting in the same location (say at the pole of a sphere) and oriented in the same direction.  We then rotate observatory $b$ by an angle $\sigma_2 - \sigma_1$ and observatory $a$ by $\sigma_1 - \pi/4$ (this is because $\sigma_1$ is measured from the \emph{bisector}).  We then rotate observatory $b$ about the $y$-axis through an angle $\beta$ and we have established our two-observatory geometry. 

In order to characterize the response of this network of observatories to the intensity and circular polarization of a stochastic gravitational wave background for each pair we must specify the angles $(\beta, \sigma_1, \sigma_2)$.  The latitude and longitude of each observatory easily allows a calculation of $\beta$ for each pair.  To determine $\sigma_1$ and $\sigma_2$ we must construct the vector tangent to the surface at the Earth at the location of each member of a pair of observatories that points along the great circle (i.e. geodesic) that connects the two.  To calculate this for each pair we used the spherical linear interpolation (SLERP) algorithm \cite{SLERP}.  The parametric equation for the geodesic that connects two points on the unit sphere,  $\hat r_1$ and $\hat r_2$, is given by 
\begin{equation}
{\rm Slerp}(\vec r_1, \vec r_2;t) = \frac{\sin[(1-t) \hat{r}_1 \cdot \hat{r}_2]}{\sin[ \hat{r}_1 \cdot \hat{r}_2]}\hat{r}_1+  \frac{\sin[t\  \hat{r}_1 \cdot \hat{r}_2]}{\sin[ \hat{r}_1 \cdot \hat{r}_2]} \hat{r}_2,
\end{equation}
where $0\leq t \leq 1$. The tangent to the sphere along the geodesic at any point is then given by $\vec T = d{\rm Slerp}/dt$.  With this tangent vector and the location of the two observatories it is straightforward to calculate $\sigma_1$ and $\sigma_2$. 

\section{Response of observations of the CMB \label{sec:CMB}}

With the SNR for CMB experiments given in Eqs.~(\ref{eq:CMBI}) and (\ref{eq:CMBV}) we estimate 
the minimum detectable signal by setting the SNR=1:
\begin{eqnarray}
I_{\rm min}({k_\ell}) &\simeq& \left[\frac{(Q_\ell^{TT})^2}{N_\ell^{TT,TT}} +\frac{(Q_\ell^{BB})^2}{N_\ell^{BB,BB}}\right]^{-1/2},\\
V_{\rm min}({k_\ell}) &\simeq& \left[\frac{(Q_\ell^{TB})^2 N_\ell^{EB,EB}-2C_\ell^{EB} C_{\ell}^{TB} N_{\ell}^{TB,EB}}{N_\ell^{EB,EB}N_{\ell}^{TB,TB}-(N_\ell^{TB,EB})^2} \right]^{-1/2},
\end{eqnarray}
where $Q^{XX'}_\ell \equiv (4\pi)^2 k_\ell^2/\tau_0 \Delta_\ell^X(k_\ell) \Delta_\ell^{X'}(k_\ell)$. 

In order to translate from the primordial power spectra to $\Omega_{(I,V)}$ today we must consider the evolution of the gravitational waves once they enter the horizon. As discussed in Ref.~\cite{Watanabe:2006qe} the spectral density of an inflationary gravitational wave background is given as
\begin{equation}
\Omega_{(I,V)}(k) = \frac{\{I(k),V(k)\} a}{12 H_0^2}k^2 \left( \frac{3 j_2(k\tau)}{k \tau}\right)^2.
\end{equation}
This expression is valid for conformal times $\tau > \tau_{eq}$ and wavenumbers $k < k_{eq}$, which is precisely the range applicable to the CMB. Evaluating it at the present-day, and after rewriting the Hubble constant $H_0^{-1} \simeq 3000/h$~Mpc in terms of the standard pivot $k_* = 0.05$~inv-Mpc, we obtain
\begin{eqnarray}
\Omega^{\rm min}_I h^2 &\simeq& 1875\, I_{\rm min}(k) \left( \frac{3 j_2(k\tau_0)}{k \tau_0} \, \frac{k}{k_*} \right)^2,\\
\Omega^{\rm min}_V h^2 &\simeq& 1875 \,V_{\rm min}(k) \left( \frac{3 j_2(k\tau_0)}{k \tau_0} \, \frac{k}{k_*} \right)^2.
\end{eqnarray}

\end{appendix}
\bibliography{CircPol.bib}

\end{document}